\documentclass{aa}
\usepackage{graphicx}
\usepackage{txfonts}
\usepackage{hyperref}
\bibpunct{(}{)}{;}{a}{}{,}

\usepackage{amsmath}    % Advanced maths commands
\usepackage{amssymb}    % Extra maths symbols
\usepackage{multicol}        % Multi-column entries in tables
\usepackage{bm}         % Bold maths symbols, including upright Greek
\usepackage{pdflscape}  % Landscape pages

\usepackage{comment}

\usepackage[T1]{fontenc}

\newcommand{\he}{HE~1143-1810}

\newcommand{\xmm}{{\em XMM--Newton}}
\newcommand{\nus}{{\em NuSTAR}}

\newcommand{\aer}[3]{$#1^{+ #2}_{- #3}$}
\newcommand{\aerm}[3]{#1^{+ #2}_{- #3}}

\newcommand{\ser}[2]{$#1 \pm #2$}

\newcommand{\serexp}[3]{($#1 \pm #2) \times 10^{#3}$}

\newcommand{\expo}[2]{$ #1 \times 10^{#2}$}
\newcommand{\expom}[2]{#1 \times 10^{#2}}
\newcommand{\tento}[1]{$10^{#1}$}
\newcommand{\tentom}[1]{10^{#1}}

\newcommand{\pers}{s$^{-1}$}
\newcommand{\chisq}{\chi^{2}}
\newcommand{\rchisq}{\chi^{2}/\textrm{dof}}
\newcommand{\dchi}{\Delta \chi^{2}}
\newcommand{\ddof}{\Delta \textrm{dof}}
\newcommand{\dcash}{\Delta C}
\newcommand{\rcash}{C/\textrm{dof}}
\newcommand{\cut}{E_{\textrm{c}}}
\newcommand{\nh}{N_{\textrm{H}}}

\newcommand{\afe}{A_{\textrm{Fe}}}
\newcommand{\mbh}{M_{\textrm{BH}}}
\newcommand{\msun}{M$_{\odot}$}

\newcommand{\ftwoten}{F_{\textrm{2--10keV}}}
\newcommand{\lbol}{L_{\textrm{bol}}}

\newcommand{\kalfa}{K$\alpha$}
\newcommand{\kbeta}{K$\beta$}
\newcommand{\fek}{Fe~K$\alpha$}

\newcommand{\ione}[2]{#1~{\sc #2}}

\newcommand{\xspec}{{\sc xspec}}
\newcommand{\xillver}{{\sc xillver}}
\newcommand{\xillvercp}{{\sc xillverCp}}
\newcommand{\xillverd}{{\sc xillverD}}
\newcommand{\relxill}{{\sc relxill}}
\newcommand{\relxilld}{{\sc relxillD}}

\newcommand{\nthcomp}{{\sc nthcomp}}

\newcommand{\zga}{{\sc zgauss}}

\newcommand{\kdblur}{{\sc kdblur}}
\newcommand{\pexrav}{{\sc pexrav}}
\newcommand{\redden}{{\sc redden}}
\newcommand{\smallbb}{{\sc smallBB}}
\newcommand{\simpl}{{\sc simpl}}
\newcommand{\borus}{{\sc borus}}
\newcommand{\relxillcp}{{\sc relxillCp}}

\newcommand{\kte}{kT_{\textrm{e}}}
\newcommand{\ktbb}{kT_{\textrm{BB}}}
\newcommand{\kteh}{kT_{\textrm{e,h}}}
\newcommand{\ktew}{kT_{\textrm{e,w}}}
\newcommand{\gammah}{\Gamma_{\textrm{h}}}
\newcommand{\gammaw}{\Gamma_{\textrm{w}}}

\newcommand{\fluxph}{photons~s$^{-1}$~cm$^{-2}$}
\newcommand{\fluxcgs}{erg~s$^{-1}$~cm$^{-2}$}
\newcommand{\lumcgs}{erg~s$^{-1}$}

\newcommand{\ovii}{O~{\sc vii}}

\newcommand{\feii}{Fe~{\sc ii}}
\newcommand{\sqcm}{cm$^{-2}$}
\newcommand{\rin}{R_{\textrm{in}}}
\newcommand{\rg}{$R_{\textrm{G}}$}

\newcommand{\refl}{\mathcal{R}}

\newcommand{\mdot}{\dot{m}}

\newcommand{\subrm}[1]{_{\textrm{#1}}}

\begin{document} 

\title{\nus/\xmm\ monitoring of the Seyfert 1 galaxy HE~1143-1810}
\subtitle{Testing the two-corona scenario}

\titlerunning{High-energy monitoring of HE 1143-1810}
\authorrunning{F. Ursini et al.}

   \author{
        F. Ursini\inst{1},
          P.-O. Petrucci\inst{2},
          S. Bianchi\inst{3},
          G. Matt\inst{3},
          R. Middei\inst{3},
          G. Marcel\inst{4},
          J. Ferreira\inst{2},
          M.~Cappi\inst{1},
          B.~De~Marco\inst{5},
          A.~De~Rosa\inst{6},
          J.~Malzac\inst{7},
          A. Marinucci\inst{8},
          G. Ponti\inst{9}
          \and
          A. Tortosa\inst{10}
          }

   \institute{
        INAF-Osservatorio di astrofisica e scienza dello spazio di Bologna, Via Piero Gobetti 93/3, 40129 Bologna, Italy\\
    \email{francesco.ursini@inaf.it}
    \and Univ. Grenoble Alpes, CNRS, IPAG, 38000 Grenoble, France %2
        \and Dipartimento di Matematica e Fisica, Universit\`a degli Studi Roma Tre, via della Vasca Navale 84, 00146 Roma, Italy %3
        \and Villanova University, Department of Physics, Villanova, PA 19085, USA %4
        \and Nicolaus Copernicus Astronomical Center, PL-00-716 Warsaw, Poland %5
        \and INAF-Istituto di Astrofisica e Planetologia Spaziali, via Fosso del Cavaliere, 00133 Roma, Italy %6
        \and IRAP, Universit\'{e} de Toulouse, CNRS, UPS, CNES, Toulouse, France %7
        \and ASI - Unità di Ricerca Scientifica, Via del Politecnico snc, I-00133 Roma, Italy %8
        \and Max-Planck-Institut f\"ur extraterrestrische Physik, Giessenbachstrasse, D-85748 Garching, Germany %9
        \and Núcleo de Astronomía de la Facultad de Ingeniería, Universidad Diego Portales, Av. Ejército Libertador 441, Santiago, Chile %10
             }

   \date{Received ...; accepted ...}

% \abstract{}{}{}{}{} 
% 5 {} token are mandatory
 
  \abstract
   {}
   {We test the two-corona accretion scenario for active galactic nuclei in the case of the `bare' Seyfert 1 galaxy \he.}
   {We perform a detailed study of the broad-band UV--X-ray spectral properties and of the short-term variability of \he. We present results of a joint \xmm\ and \nus\ monitoring of the source, consisting of $5 \times 20$ ks observations, each separated by 2 days, performed in December 2017.}
   {The source is variable in flux among the different observations, and a correlation is observed between the UV and X-ray emission. Moderate spectral variability is observed in the soft band. The time-averaged X-ray spectrum exhibits a cut-off at $\sim 100$ keV consistent with thermal Comptonization. 
        We detect an iron \kalfa\ line consistent with being constant during the campaign and originating from a mildly ionized medium. The line is accompanied by a moderate, ionized reflection component. A soft excess is clearly present below 2 keV and is well described by thermal Comptonization in a `warm' corona with a temperature of $\sim 0.5$ keV and a Thomson optical depth of $\sim 17-18$. For the hot hard X-ray emitting corona, we obtain a temperature of $\sim 20$ keV and an optical depth of $\sim 4$ assuming a spherical geometry. 
        A fit assuming a jet-emitting disc (JED) for the hot corona also provides a nice description of the broad-band spectrum. 
        In this case, the data are consistent with an accretion rate varying between $\sim 0.7$ and $\sim 0.9$ in Eddington units and a transition between the outer standard disc and the inner JED at $\sim 20$ gravitational radii.}
   {
                The broad-band high-energy data agree with an accretion flow model consisting of two phases: an outer standard accretion disc with a warm upper layer, responsible for the optical--UV emission and the soft X-ray excess, and an inner slim JED playing the role of a hard X-ray emitting hot corona. 
}


   \maketitle
%
%-------------------------------------------------------------------

\section{Introduction}
The X-ray emission of active galactic nuclei (AGNs) is believed to originate from thermal Comptonization of optical--UV photons, emitted by the accretion disc, in a hot corona  \cite[see e.g.][]{haardt&maraschi1991,hmg1994,hmg1997}. 
This physical mechanism is able to explain the observed power-law shape of the X-ray emission. Moreover, the primary continuum often exhibits a high-energy cut-off at around 100-150 keV, which is interpreted as the roll-over of thermal Comptonization due to the finite coronal temperature. This feature has been observed in a number of sources \cite[see e.g.][]{zdziarski2000,perola2002,dadina2008,bat70,malizia2014,lubinski2016}, in particular from high-sensitivity measurements enabled by \nus\ \cite[]{fabian2015,tortosa2018}. 
In addition to the primary emission, other spectral components are often observed, such as a hump at 20-30 keV interpreted as Compton reflection from the accretion disc or more distant material \cite[see e.g.][]{george&fabian1991,MPP1991}, and a soft excess below 1-2 keV with a steep rising shape \cite[see e.g.][]{caixa1}. 
Currently, the most debated models for the origin of the soft excess are 
ionized reflection \cite[see e.g.][]{crummy2006,ponti2006,walton2013,jiang2018,garcia2019} and `warm' Comptonization \cite[see e.g.][]{mag_5548,mehdipour2011509,done2012SE,pop2013mrk509,rozenn2014mrk509SE,matt2014ark120,middei2018,porquet2018,cheeses,3c382,4593_2}. In the latter hypothesis the optical--UV emission and soft X-ray excess could originate from the upper layer of the accretion disc, consisting of a warm ($\kte\sim 1$ keV) optically thick ($\tau \sim 10-20$) slab-like corona \cite[]{rozanska2015,cheeses}.

In this paper we investigate the properties of the hot corona and the physical origin of the soft excess in the Seyfert 1 galaxy \he\ \cite[$z=0.0328$,][]{z_he} through a joint \xmm\ and \nus\ monitoring programme carried out in 2017. 
A previous \xmm\ observation of the source in 2004 revealed the presence of a significant soft excess and of a narrow \fek\ emission line at 6.4 keV \cite[]{cardaci2011}, with ambiguous evidence for a relativistically broadened component \cite[]{nandra2007pexmon,bhayani&nandra2011}. 

The structure of the paper is as follows. In Section \ref{sec:obs} we describe
the observations and data reduction. In Section \ref{sec:timing} we discuss the timing properties. In Section \ref{sec:analysis} we present the analysis of the \xmm\ and \nus\ spectra. In Section \ref{sec:discussion} we discuss the results and our conclusions.

%--------------------------------------------------------------------
\section{Observations and data reduction}\label{sec:obs}
\he\ was observed five times simultaneously by \xmm\ \cite[]{xmm} and \nus\ \cite[]{harrison2013nustar} in December 2017. The pointings had a net exposure of ${\sim} 20$ ks each, and were separated by 2 days from each other. The log of the data sets is listed in Table \ref{tab:log}. 

\begin{table}[]
        \setlength{\tabcolsep}{5pt}
        \small
        \caption{Logs of the \xmm\ and \nus\ observations of \he. \label{tab:log}}
        \centering
        \begin{tabular}{ c c c c c} 
                \hline \hline 
                Obs. & Satellites &  Obs. Id. & Start time (\textsc{utc})  & Net exp.\\ 
                & & & yyyy-mm-dd & (ks)  \\ \hline 
                1 
                & \xmm  & 0795580101  & 2017-12-16 & 23 \\ 
                & \nus  
                & 60302002002 &  & 21 \\
                \hline 
                2
                & \xmm  & 0795580201  & 2017-12-18 & 20 \\ 
                & \nus  
                & 60302002004 &  & 21 \\
                \hline 
                3
                & \xmm  & 0795583101  & 2017-12-20 & 20 \\ 
                & \nus  
                & 60302002006 &  & 23 \\
                \hline 
                4
                & \xmm  & 0795580401  & 2017-12-22 & 19 \\ 
                & \nus  
                & 60302002008 &  & 21 \\
                \hline          
                5
                & \xmm  & 0795580501  & 2017-12-24 & 20 \\ 
                & \nus  
                & 60302002010 &  & 22 \\
                \hline          
        \end{tabular}
\end{table} 

\xmm\ observed the source with the optical monitor \cite[OM;][]{OM}, the EPIC cameras \cite[][]{pn,MOS}, and the Reflection Grating Spectrometer \cite[RGS;][]{RGS}.
The data were processed using the \xmm\ Science Analysis System (\textsc{sas} v18).
The OM photometric filters were operated in the image mode; the images were taken with the U, UVW1, UVM2, and UVW2 filters, with an exposure time of 5 ks each. The OM data were processed with the \textsc{sas} pipeline \textsc{omichain}, and converted into data suitable for the spectral analysis using the \textsc{sas} task \textsc{om2pha}.
The EPIC-pn instrument operated in the Small Window mode, with the thick filter applied, while the EPIC-MOS instruments operated in the Timing mode. 
The spectral analysis is based on pn data, because of the higher effective area compared with MOS and to avoid cross-calibration uncertainties. 
The data show no significant pile-up as indicated by the \textsc{sas} task \textsc{epatplot}. 
Source extraction radii and screening for high-background intervals were determined through an iterative process that maximizes the signal-to-noise ratio \cite[][]{pico2004}. The background was extracted from circular regions with a radius of 50 arcsec, while the source extraction radii were allowed to be in the range 20--40 arcsec; the best extraction radius was found to be 40 arcsec for every iteration. The light curves were corrected for instrumental effects and were background-subtracted using the \textsc{sas} task \textsc{epiclcorr}. The EPIC-pn spectra were grouped such that each spectral bin contained at least 30 counts, and without oversampling the spectral resolution by a factor greater than 3. Finally, the RGS data were extracted using the standard \textsc{sas} task \textsc{rgsproc}. The spectra from the two detectors RGS1 and RGS2 were not combined.

The \nus\ data were reduced using the standard pipeline (\textsc{nupipeline}) in the \nus\ Data Analysis Software (\textsc{nustardas}, v1.9.3), using calibration files from \nus\ {\sc caldb} v20171002. Spectra and light curves were extracted using the standard tool {\sc nuproducts} for each of the two hard X-ray detectors aboard \nus, located inside the corresponding focal plane modules A and B (FPMA and FPMB). The source data were extracted from circular regions with a radius of 75 arcsec, and the background was extracted from a blank area close to the source. The spectra were binned to have a signal-to-noise ratio greater than 5 in each spectral channel, and without  oversampling the instrumental resolution by a factor greater than 2.5. The spectra from FPMA and FPMB were analysed jointly, but were not combined.

\begin{figure*}[h!] 
        \includegraphics[width=2\columnwidth]{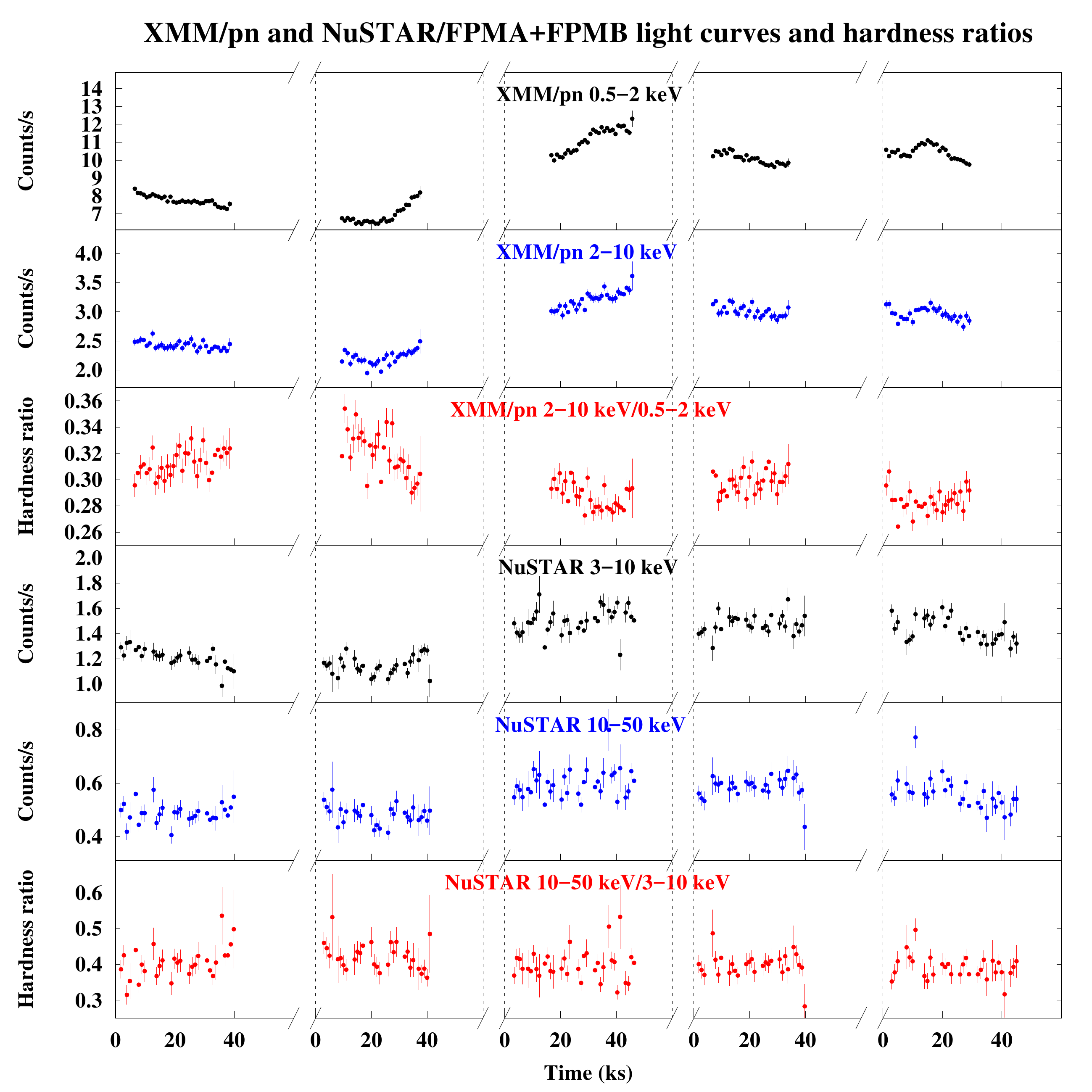}
        \caption{\label{fig:lc} Light curves of the five joint \xmm\ and \nus\ observations of \he. The exposures are spaced by 2 d. Time bins of 1 ks are used. Top panel: \xmm/pn count rate light curve in the 0.5--2 keV band. Second panel: \xmm/pn count rate light curve in the 2--10 keV band. Third panel: \xmm/pn hardness ratio (2--10/0.5--2 keV) light curve. Fourth panel: \nus\ count rate light curve in the 3--10 keV band (FPMA and FPMB data are co-added). Fifth panel: \nus\ count rate light curve in the 10--50 keV band. Bottom panel: \nus\ hardness ratio (10--50/3--10 keV) light curve.}
\end{figure*}

\section{Timing properties}\label{sec:timing}
The light curves of \he\ with \xmm/pn and \nus, in different energy ranges, are plotted in Fig. \ref{fig:lc}. The source exhibits a moderate flux variability between different observations, up to  a factor of 1.7 in the 0.5--2 keV band. In Fig. \ref{fig:lc} we also plot the pn (2--10 keV)/(0.5--2 keV) hardness ratios and the \nus\ (10--50 keV)/(3--10 keV) hardness ratios. The soft band (0.5--10 keV) displays the most significant spectral variability between different observations; however, it is no greater than 20\% in terms of the pn hardness ratio.

In order to characterize the flux variability, we computed the normalized excess variance \cite[e.g.][]{nandra1997,vaughan2003,caixa3}, defined as
\begin{equation}
\sigma^2_{\rm{rms}}=\frac{1}{N\mu^2}\sum_{i=1}^{N} \left[ (X_i-\mu)^2-\sigma_i^2 \right]
,\end{equation}
where $N$ is the number of time bins in the light curve, $\mu$ is the unweighted mean of the count rate within that segment, $X_i$ is the count rate, and $\sigma_i^2$ is the associated uncertainty. Following \cite{caixa3}, we computed the normalized excess variance in the 2--10 keV band for all the observations of our campaign; the light curves were calculated using time bins of 250 s and selecting segments of 20 ks. 
We also included the light curve of the 2004 \xmm\ observation \citep{cardaci2011}. We obtained $\sigma^2_{\rm{rms}}= \expom{\aerm{7}{6}{4}}{-4}$. 
Then, assuming the correlation between $\sigma_{\rm{rms}}^2$ and the black hole mass $\mbh$ reported by \cite{caixa3}, we estimate $\mbh = \expom{\aerm{7}{5}{4}}{7}$ \msun. 
From the properties of the H $\beta$ emission line as reported by \cite{winkler1992} and \cite{marziani2003}, we derive a black hole mass of \expo{3-4}{7} \msun\ applying the virial mass estimators of \cite{shen&liu2012} and \cite{ho&kim2015}. These results are consistent, and hereafter we assume a mass of \expo{4}{7} \msun.

\begin{figure} 
        \includegraphics[width=\columnwidth]{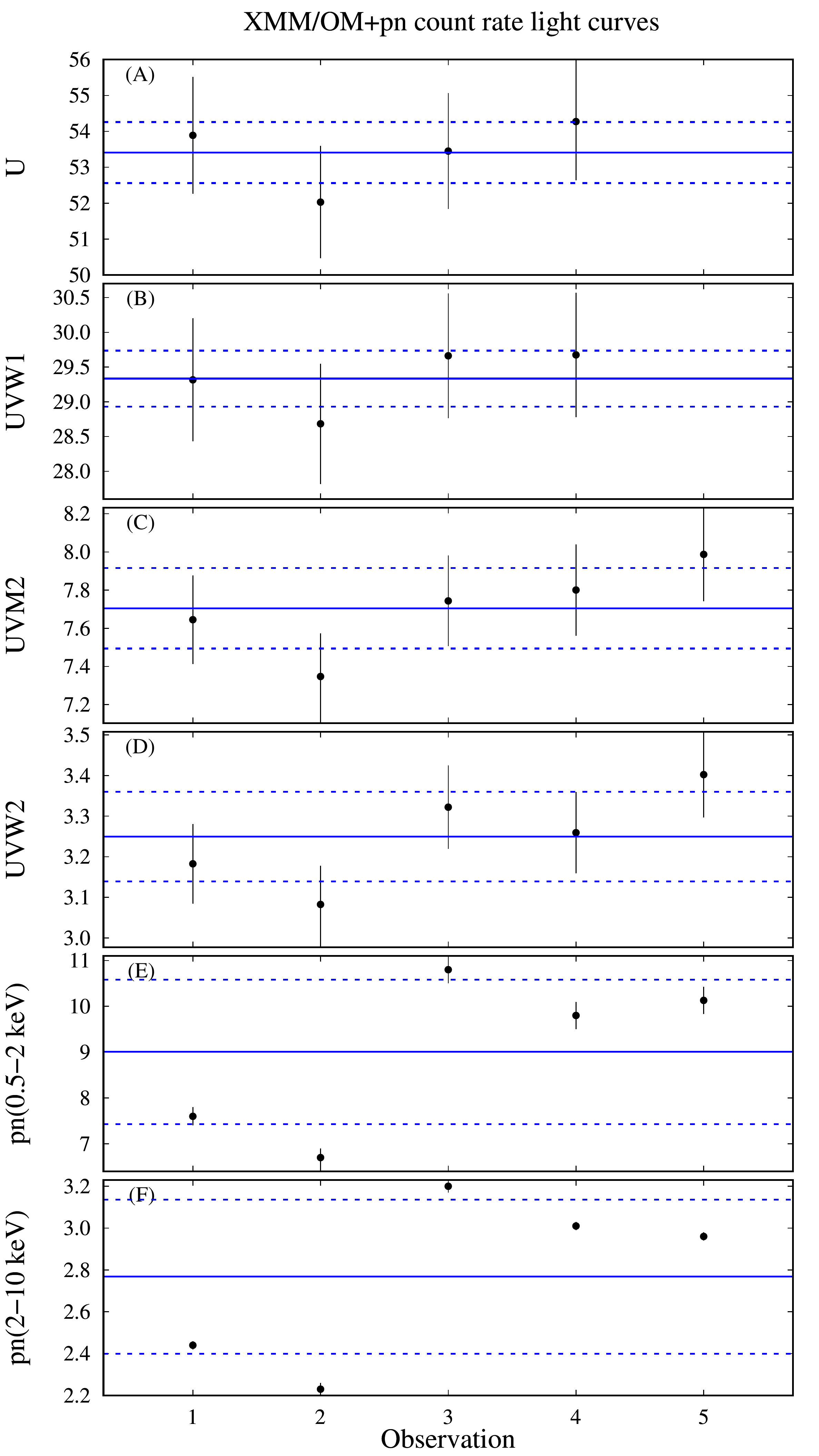}
        \caption{\label{fig:om_lc} Light curves of each of the four \xmm/OM photometric filters: U (panel A), UVW1 (panel B), UVM2 (panel C), UVW2 (panel D); and light curves, averaged over each observation, of \xmm/pn in the bands 0.5--2 keV (panel E) and 2--10 keV (panel F). The U and UVW1 filters were not available during Obs. 5. The blue solid lines represent the mean value of the count rate over the five observations, while the blue dashed lines represent the standard deviation (i.e. the root mean square of the deviations from the mean). }
\end{figure}

\begin{figure} 
        \includegraphics[width=\columnwidth]{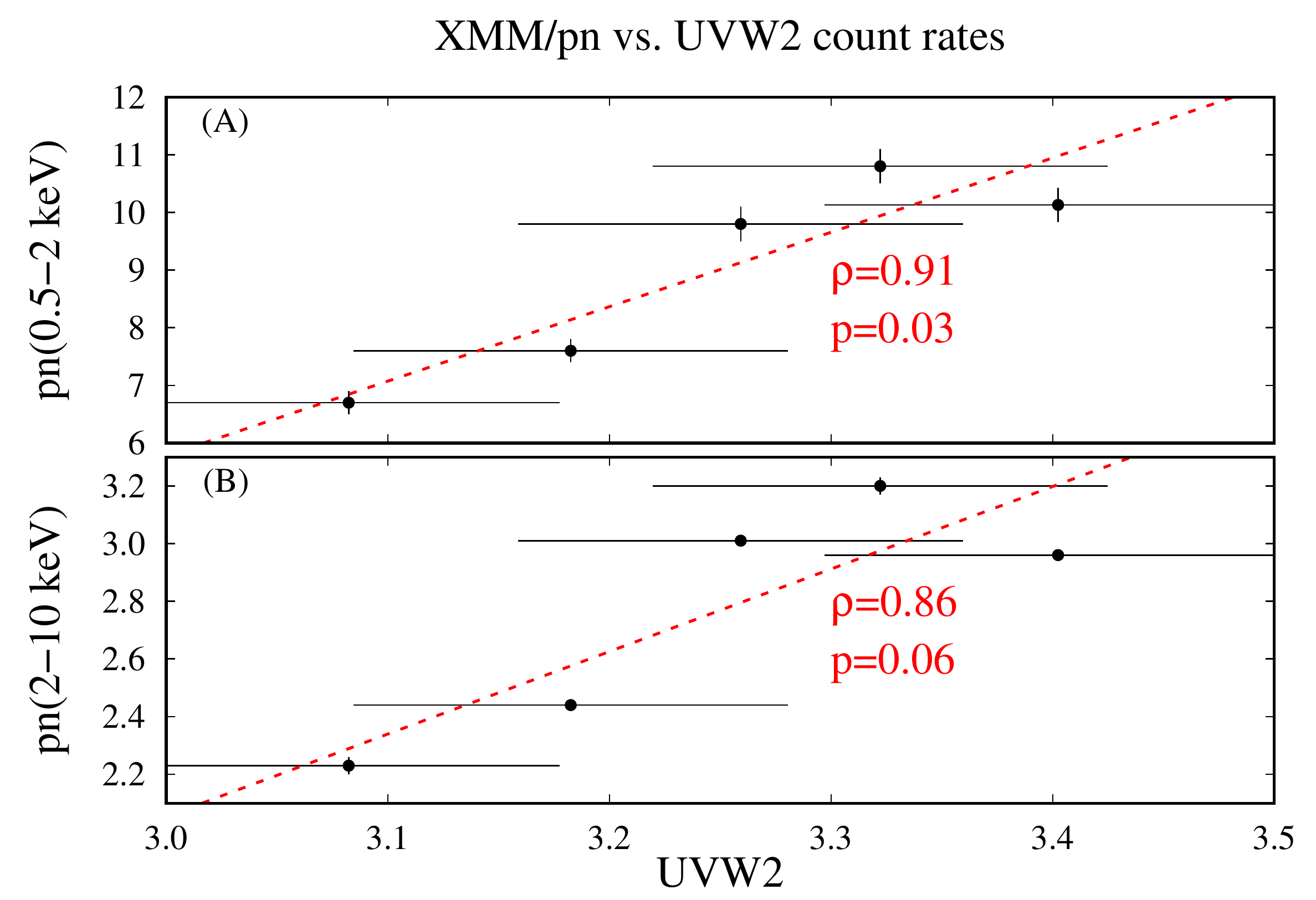}
        \caption{\label{fig:pn_vs_om} \xmm/pn count rate, averaged over each observation, in the 0.5--2 keV band (panel A) and in the 2--10 keV band (panel B) plotted against the OM/UVW2 count rate. The red dashed lines represent linear fits to the data.}
\end{figure}

In Fig. \ref{fig:om_lc} we plot the light curves of the four \xmm/OM filters, and the \xmm/pn average count rate measured for each observation in the bands 0.5--2 keV and 2--10 keV. 
The UVW2 filter exhibits marginal variability. In Fig. \ref{fig:pn_vs_om} we plot the \xmm/pn average count rates for each observation versus the OM/UVW2 count rate. 
The correlation between UVW2 and 0.5--2 keV band has a Pearson's correlation coefficient of 0.91, with a null hypothesis probability of 0.03. For the 2--10 keV band, the Pearson's coefficient is 0.86 and the null hypothesis probability is 0.06. 
Such correlations, albeit not highly significant, indicate a trend of a higher X-ray flux with increasing UV flux.

\section{Spectral analysis}\label{sec:analysis}
We performed the spectral analysis with the \xspec\ v12.10 package \cite[][]{arnaud1996}. 
The RGS spectra were not rebinned and were analysed using the $C$-statistic \cite[][]{cstat} to exploit the high spectral resolution of the gratings in the 0.3--2 keV band. 
Broad-band fits (UV to X-ray, 0.3--79 keV)  were performed on the rebinned pn and \nus\ spectra plus the OM photometric data, using the $\chisq$ minimization technique. All errors are quoted at the 90\% confidence level ($\dcash = 2.71$ or $\dchi = 2.71$) for one interesting parameter. In our fits we always included neutral absorption ({\sc phabs} model in {\sc xspec}) from Galactic hydrogen with column density $\nh = 3.47 \times 10^{20}$ \sqcm\ \cite[][]{kalberla2005}. When using optical--UV data, we also included interstellar extinction (\redden\ model in \xspec) with $E(B-V)=0.035$ \cite[]{schlafly2011}.
We assumed the element abundances of \cite{lodders2003} and the photoelectric absorption  cross sections of \cite{vern}.

\begin{figure} 
        \includegraphics[width=\columnwidth]{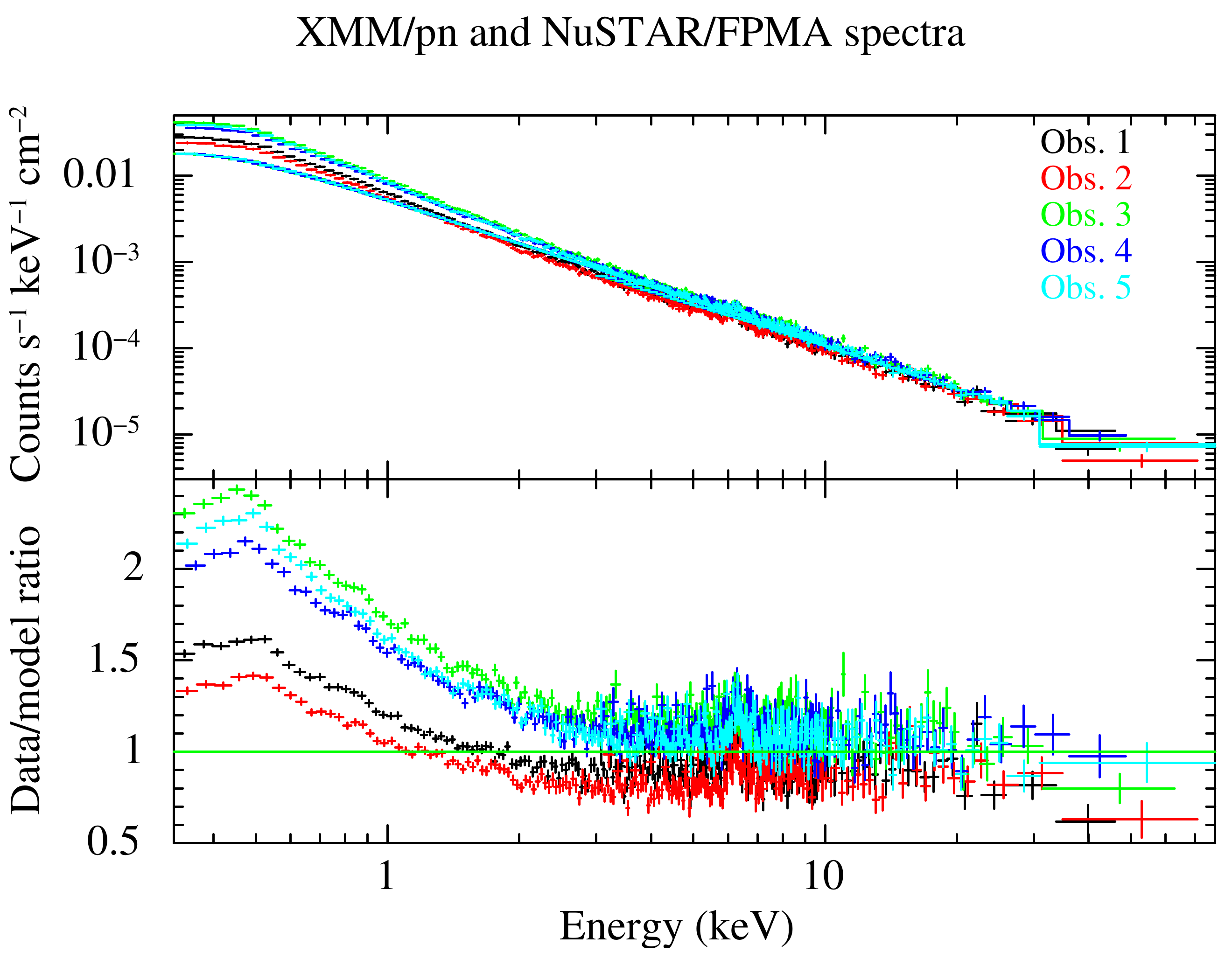}
        \caption{\label{spectra} Upper panel:  \xmm/pn and \nus/FPMA  spectra of \he; these five spectra were fitted with a single power law in the 3--79 keV band. Lower panel: Ratio of the spectra to the 3--79 keV power law. The data were binned for plotting purposes.}
\end{figure}

In Fig. \ref{spectra} we show the \xmm/pn and \nus/FPMA  spectra, fitted in the 3--79 keV band with a simple power law with parameters tied between the different detectors and observations. The \nus\ spectra exhibit a curvature above $\sim 20$ keV, while the extrapolation of pn data below 3 keV shows a significant soft excess. The \fek\ emission line is also visible in the residuals.
As already reported in \xmm\ and \nus\ simultaneous observations of other sources, the pn spectra are flatter than the \nus\ spectra in the common bandpass 3--10 keV, with a difference in photon index of ${\sim}0.1$ \cite[e.g.][]{cappi_5548,fuerst2016,middei2018,ponti2018}. In some cases the largest discrepancy was found in the 3--5 keV band, where \nus\ measures a higher flux \cite[e.g.][]{fuerst2016,ponti2018}. However, in our case the spectral discrepancy does not depend on the energy band.
We discuss this issue in more detail in Appendix \ref{app:pn-nustar}.
To account for this discrepancy, we included in the fits involving both \xmm\ and \nus\ a cross-calibration function 
in the form $\textrm{const}\times E^{\Delta \Gamma}$, where $\Delta \Gamma$ is the discrepancy in photon index between pn and \nus\ \citep[][]{ingram2017}. We fixed $\Delta \Gamma$ at zero for both \nus\ modules and left it free for pn (but tied between the different observations, see Appendix \ref{app:pn-nustar}). 
The values of the photon index and flux reported in the following (Sect. \ref{subsec:refl}, \ref{subsec:fit}, and \ref{subsec:jed}) are those measured by \nus, unless otherwise stated. 
The FPMA and FPMB modules are in very good agreement with each other, with a cross-calibration factor of \ser{1.02}{0.01}. 

The analysis of RGS data is given in Appendix \ref{app:rgs}

\subsection{High-energy turnover}\label{subsec:cutoff}
To constrain the presence of a high-energy cut-off, we fitted the time-averaged \nus\ spectra (FPMA and FPMB), co-adding the data from the five observations. 
We ignored the 5--8 keV band to avoid the contribution from the \fek\ line at 6.4 keV.

\begin{figure} 
        \includegraphics[width=\columnwidth]{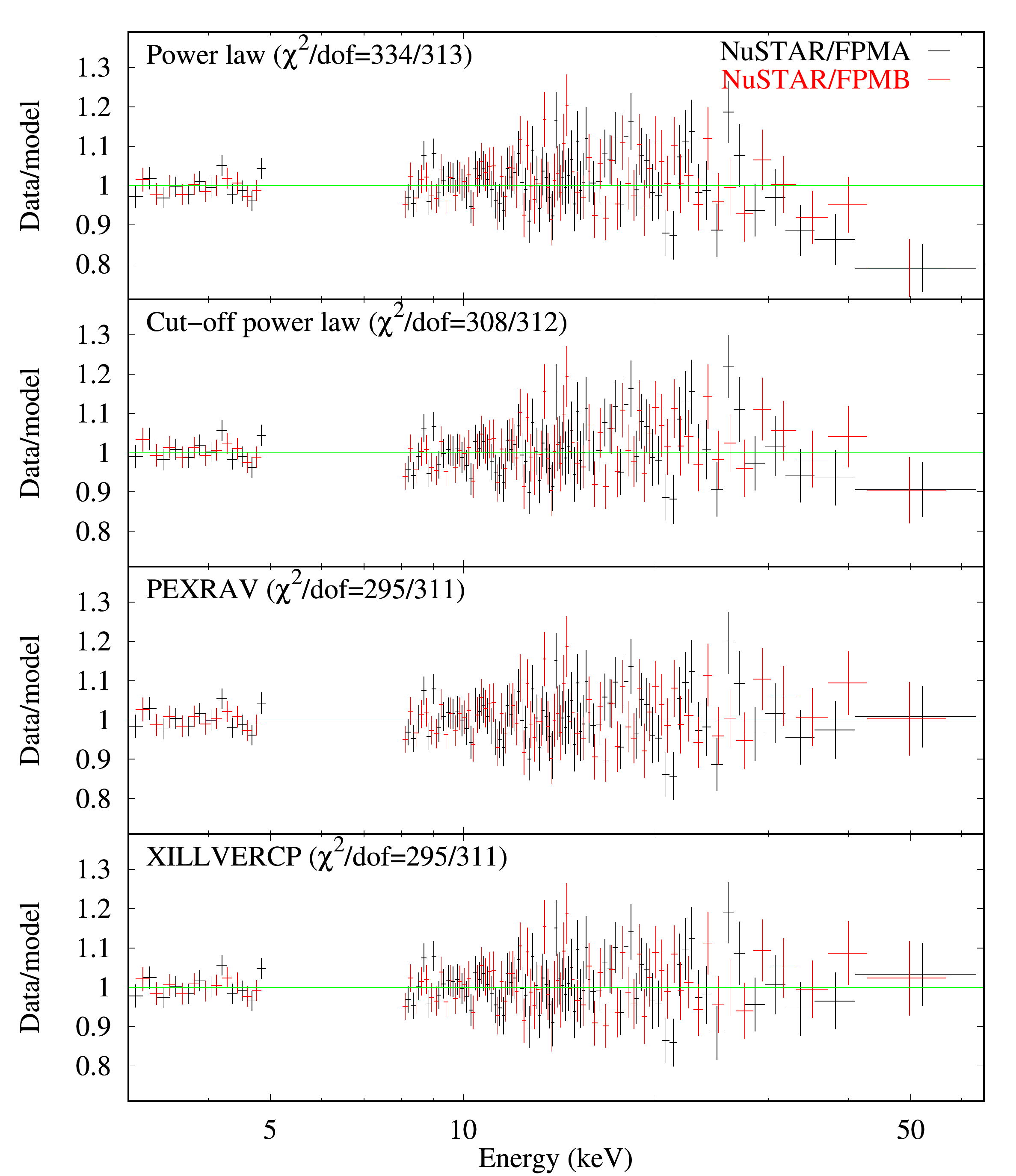}
        \caption{\label{fig:ratios} Residuals of the fits of the time-averaged \nus\ spectra with different models. Upper panel: Simple power law. Second panel: Power law with exponential cut-off . Third panel: Power law plus reflection (\pexrav). Lower panel: Thermal Comptonization model plus reflection (\xillvercp). The data were binned for plotting purposes.}
\end{figure}

Starting with a simple power law, we obtained a fit with $\rchisq=334/313$ 
that clearly indicates a turnover at around 30 keV (Fig. \ref{fig:ratios}, upper panel). Then we included an exponential high-energy cut-off, finding an improved fit with $\rchisq=308/312$ ($\dchi/\ddof=-26/-1$). The cut-off energy is found to be \aer{150}{110}{50} keV. However, despite the improvement, the fit with a power law with exponential cut-off still leaves significant residuals in the high-energy band (Fig. \ref{fig:ratios}, second panel).

We then included Compton reflection, replacing the cut-off power law with \pexrav\ \cite[]{pexrav}. This model includes Compton reflection off a neutral slab of infinite column density. We fixed the inclination angle at 30 deg, since the fit was not sensitive to this parameter. We assumed solar abundances. We left free the reflection fraction $\mathcal{R} = \Omega / 2 \pi$, where $\Omega$ is the solid angle subtended by the reflector. We obtained a good fit with $\rchisq=295/311$ ($\dchi/\ddof=-13/-1$) and without prominent residuals (Fig. \ref{fig:ratios}, third panel). In this case we obtained $\cut=\aerm{100}{50}{20}$ keV and $\mathcal{R} = \aerm{0.17}{0.09}{0.08}$.  

Finally, we tested a thermal Comptonization model. We fitted the spectra with the model \xillvercp, that includes the Comptonization model \nthcomp\ \cite[]{nthcomp1,nthcomp2} plus ionized reflection with the code \xillver\ \citep{xillver1,xillver2,xillver3}.
We left free the photon index $\Gamma$ of the asymptotic power law, the electron temperature $\kte$, and the reflection fraction. We fixed the inclination angle at 30 deg, the ionization parameter at $\log \xi =0$, and the iron abundance at the solar value.
We obtained a good fit (i.e. equivalent to the \pexrav\ fit) with $\rchisq=295/311$ (Fig. \ref{fig:ratios}, last panel). The electron temperature was found to be \aer{20}{5}{3} keV, while we only had an upper limit of 0.13 to the reflection fraction.

We conclude that the spectral turnover is nicely described by either a power law with exponential cut-off plus modest reflection or by a thermal Comptonization model. 
In the following we analyse the \fek\ line (Sect. \ref{subsec:line}) and investigate the presence of a reflection component including pn data as well (Sect. \ref{subsec:refl}).

\subsection{The \fek\ line}\label{subsec:line}
\begin{figure} 
        \includegraphics[width=\columnwidth]{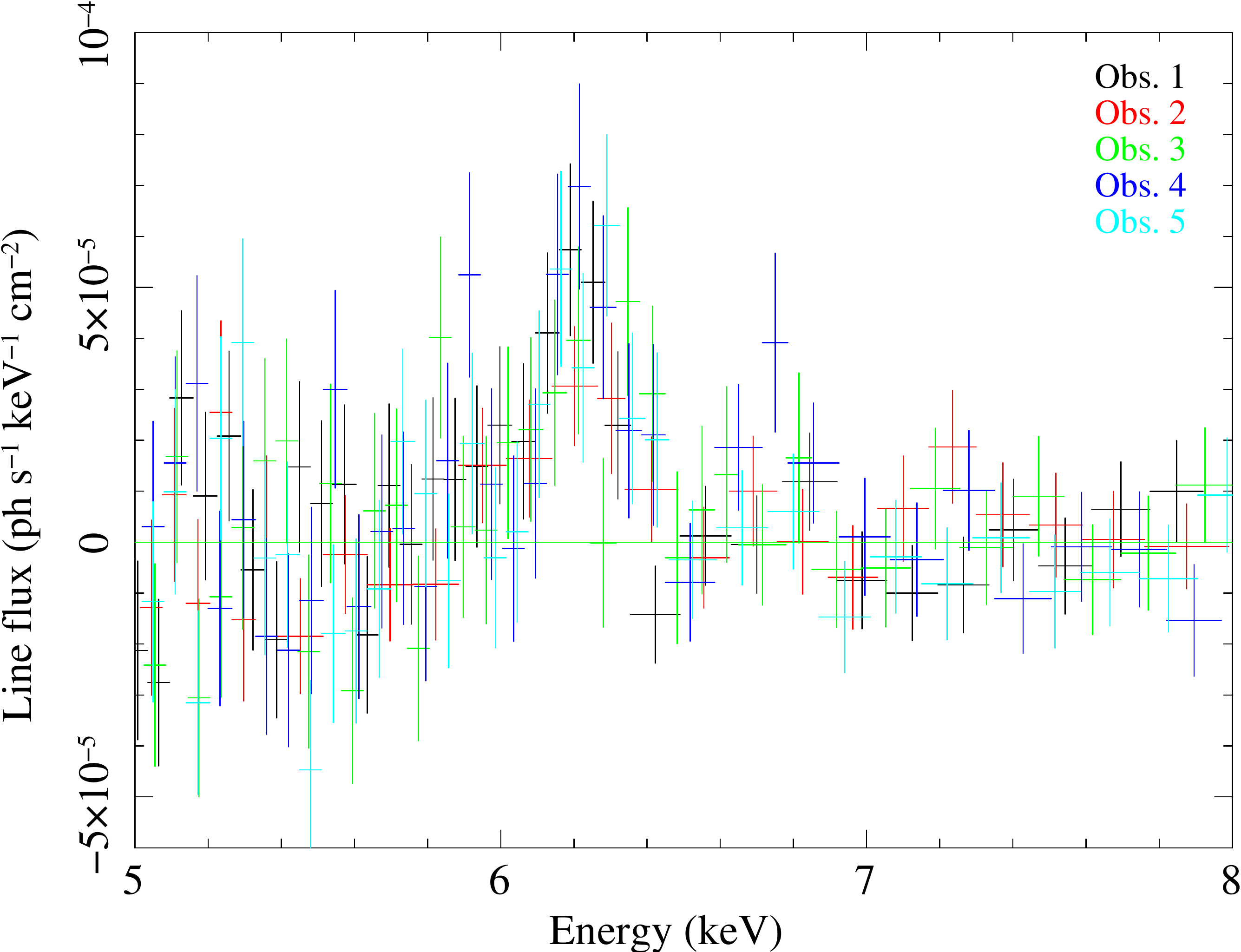}
        \caption{\label{fig:resline} \fek\ line profile from \xmm/pn data (observer's frame). The plot shows the residuals of a simple power law fit performed in the 3--10 keV band.}
\end{figure}

To investigate the shape and variability of the \fek\ line at 6.4 keV, we focused on \xmm/pn data between 3 and 10 keV because of the better energy resolution and throughput compared with \nus\ in that energy band. 
In Fig. \ref{fig:resline} we plot the profile of the \fek\ line from pn data.
We simultaneously fitted the five pn spectra with a model consisting of a variable power law plus a Gaussian line.
We first assumed an intrinsically narrow single emission line (i.e. the intrinsic width $\sigma$ was fixed at zero) with a constant flux among the different observations. We found $\rchisq= 505/511$, with positive residuals around 6.7 keV (observer's frame). We thus included a second, narrow Gaussian line, keeping its flux tied among the different observations, obtaining $\rchisq= 496/509$ ($\dchi/\ddof = -9/{-}2$).  The rest-frame energy of the second line was found to be \ser{7.00}{0.07} keV, consistent with both the \ione{Fe}{xxvi} \kalfa\ line at 6.966 keV and the neutral Fe \kbeta\ line at 7.056 keV. This line is weak in any case; it has a flux of \serexp{3.4}{1.8}{-6} photons \sqcm\ \pers\ and an equivalent width (EW) 
in the range 5--30 eV.

\begin{figure} 
        \includegraphics[width=\columnwidth]{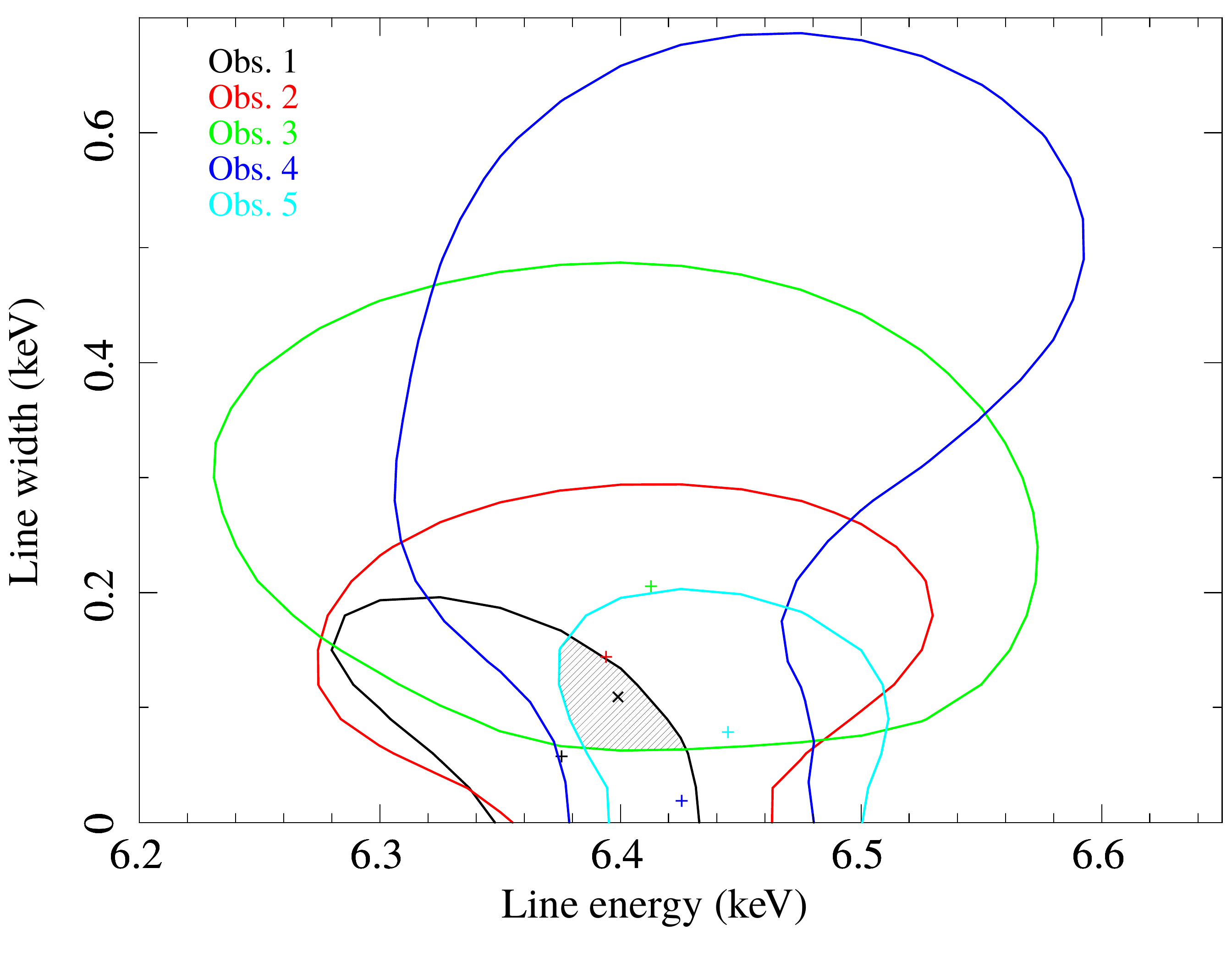}
        \caption{\label{fig:cont_en_sigma} Contour plots of the \fek\ line intrinsic width vs. rest-frame energy at the 90\% confidence level. The shaded area represents the intersection of the contours, while the black cross corresponds to the average values ($E=6.40$ keV, $\sigma=0.11$ keV).}
\end{figure}

\begin{figure}
        \includegraphics[width=\columnwidth]{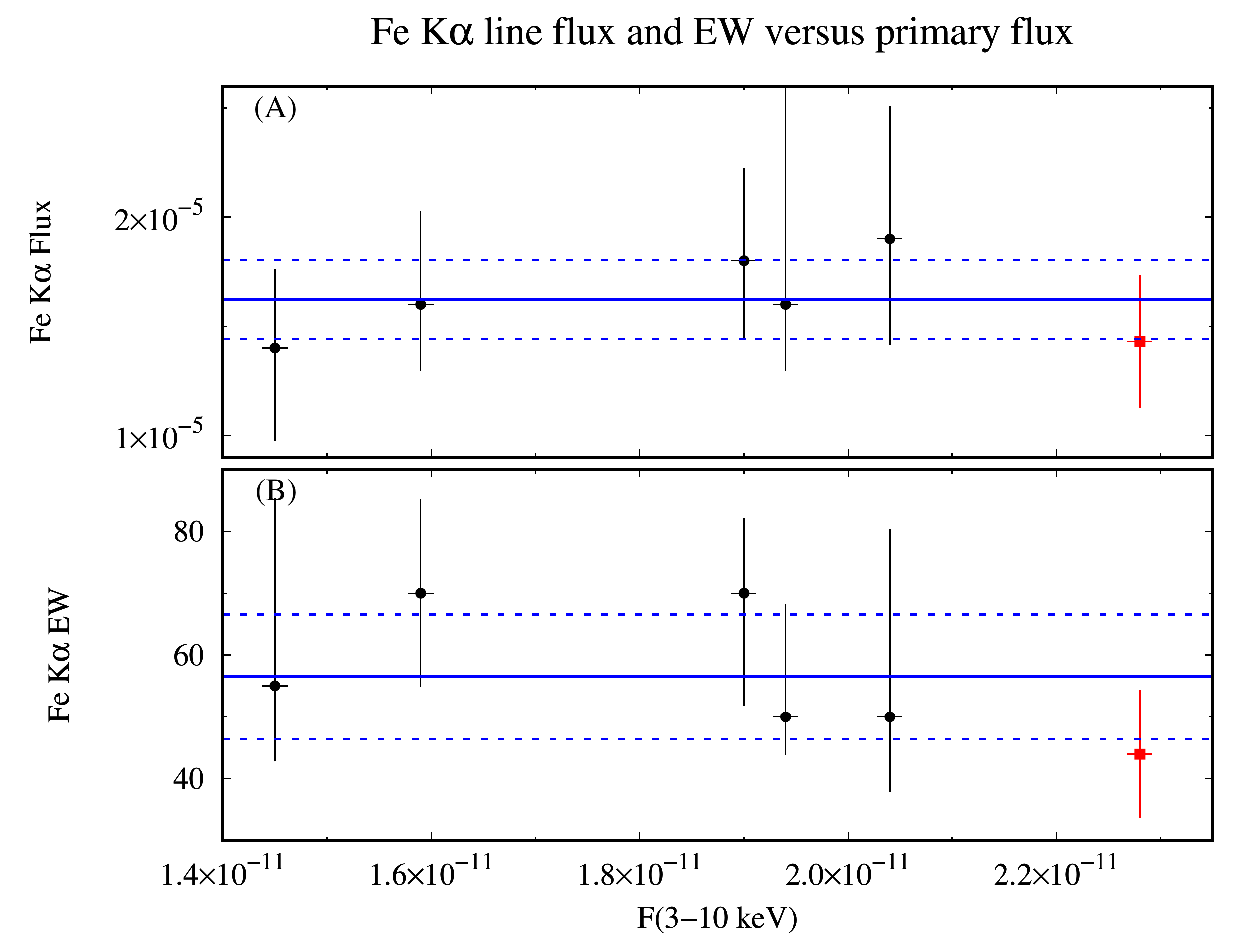}
        \caption{Parameters of the \fek\ line plotted against the primary flux in the 3--10 keV band. The red square corresponds to the 2004 \xmm\ observation.
                Panel (A): Line flux in units of 
                photons \sqcm~\pers. 
                Panel (B): Line equivalent width in units of eV. 
                Error bars denote the $1\sigma$ uncertainty. The blue solid lines represent the mean value for each parameter during the campaign, while the blue dashed lines represent the standard deviation. }
        \label{fig:line_flux}
\end{figure}

Next we tested for the variability of the neutral \fek\ line as follows. 
We first left the line flux free to vary among the observations, and found no strong improvement ($\rchisq= 490/505$, i.e. $\dchi/\ddof = -6/{-}4$). Then, we left the energy free to vary, and obtained an equivalent fit ($\rchisq=486/501$, i.e. $\dchi/\ddof = -4/{-}4$). Then we left the intrinsic width free, but tied between the observations, and found a minor improvement ($\rchisq=481/500$, i.e. $\dchi/\ddof = -5/{-1}$). Finally, we left the line width free to vary among the different observations, but found no strong improvement ($\rchisq=475/496$, i.e. $\dchi/\ddof = -6/{-}4$).
The contours of the line intrinsic width versus rest-frame energy for the different observations are plotted in Fig. \ref{fig:cont_en_sigma}; all 
the parameters are summarized in Table \ref{tab:line}.
We also reanalysed the archival 2004 \xmm/pn observation, as above, to investigate the \fek\ line variability on longer timescales. 
The parameters, which are listed in Table \ref{tab:line}, are consistent with those found by \cite{cardaci2011} and agree within the errors with those of the 2017 campaign.
We conclude that the line is consistent with being constant in flux during the 2017 monitoring, with no significant variations since 2004 (see Fig. \ref{fig:line_flux}). 

The line has an intrinsic width of $\sim 0.11$ keV and a rest-frame energy of $\sim 6.40$ keV.
We then tested a model in which the Gaussian line is broadened by relativistic effects in the inner region of the accretion disc. We used a narrow Gaussian line ($\sigma=0$)  convolved with \kdblur\ in \xspec. 
We left the inner disc radius $\rin$ free to vary among the different observations, and fixed the outer disc radius to 400 \rg\ (the fit being insensitive to this parameter). The disc inclination was fixed at 30 deg, with no improvement by leaving it free to vary. We obtained a fit with $\rchisq=505/496$,  worse than the fit with a free $\sigma$ ($\dchi=+30$), and lower limits on the inner disc radius of about 100 \rg.
The moderate broadening and the possible presence of a 7 keV feature could indicate that the line is produced in a mildly ionized medium, possibly being a blend of the \kalfa\ emission from intermediate ions of iron \cite[e.g.  \ione{Fe}{xvii}--\ione{Fe}{xx};][]{xillver2,xillver3}. We  discuss this point in the following analysis.

\begin{table*}
        \begin{center}
                \caption{
                        Properties of the \fek\ emission line: $E$ is the energy of the line (rest-frame) in keV, $\sigma$ is the intrinsic line width in keV, the flux is in units of $10^{-5}$ photons \sqcm\ \pers, and EW is the equivalent width in eV. The second column is relative to the 2004 \xmm\ observation.
                        \label{tab:line}}
                \begin{tabular}{l c c c c c c}
                        \hline 
                        &2004 &Obs. 1&Obs. 2&Obs. 3&Obs. 4&Obs. 5\\
                        \hline
                        $E$&\ser{6.40}{0.05}&\aer{6.38}{0.05}{0.07}&\ser{6.40}{0.10}&\aer{6.41}{0.12}{0.13}&\aer{6.43}{0.04}{0.05}&\ser{6.45}{0.05}\\
                        $\sigma$&$<0.12$&$<0.16$&$<0.25$&\aer{0.2}{0.2}{0.1}&$<0.7$&$<0.2$\\
                        flux& \ser{1.4}{0.5}&\aer{1.6}{0.7}{0.5}&\aer{1.4}{0.7}{0.6}&\aer{1.9}{1.0}{0.8}&\aer{1.6}{1.8}{0.5}&\aer{1.8}{0.7}{0.6}\\
                        EW &\ser{44}{17}&\ser{70}{25}&\aer{55}{50}{20}&\aer{50}{50}{20}&\aer{50}{30}{10}&\aer{70}{20}{30}\\  
                        \hline
                \end{tabular}
        \end{center}
\end{table*}

\subsection{Reflection component}\label{subsec:refl}
We tested the presence of a Compton reflection hump 
fitting simultaneously the pn and \nus\ data in the 3--79 keV band. 
First, we fitted the five observations with a simple model consisting of a power law with an exponential cut-off plus two Gaussian lines. The photon index, cut-off energy, and normalization of the power law were free to vary among the different observations.
Since the neutral \fek\ line is consistent with being constant (Sect. \ref{subsec:line}), we kept the energy, width, and flux of the Gaussian lines tied among the observations. 
We fixed the width of the 7 keV line at zero.
We found a good fit with $\rchisq=1584/1646$.
We then replaced the cut-off power law with \pexrav. We fixed the inclination angle at 30 deg since the fit was not sensitive to this parameter. We assumed a constant reflection fraction $\mathcal{R}$. We found $\rchisq=1577/1645$ ($\dchi/\ddof=-7/-1$) and $\mathcal{R} = \aerm{0.11}{0.08}{0.07}$, and no improvement by leaving $\refl$ free to vary among the observations. 

Then we replaced \pexrav\ and the Gaussian line with 
\xillver\ (which self-consistently incorporates fluorescence lines and the Compton hump) assuming illumination from a cut-off power law spectrum. We fixed the inclination angle at 30 deg, leaving the iron abundance $\afe$ and the ionization parameter $\log \xi$ as free parameters, but constant among the different observations. We found a good fit with $\rchisq=1584/1648$ 
and $\mathcal{R} = \aerm{0.18}{0.06}{0.05}$, $\afe=\aerm{3}{4}{2}$, and $\log \xi < 2.1$ \lumcgs\ cm.
Next we replaced \xillver\ with \relxill, which describes relativistically blurred reflection off an ionized accretion disc
\cite[][]{relxill,dauser2016}. 
The iron abundance and the ionization parameter were free and tied, as in the \xillver\ fit, while we left 
the inner disc radius free to vary among the different observations. Furthermore, we left  the inclination $i$ free.
We found no improvement ($\rchisq=1582/1642$, i.e. $\dchi/\ddof = -2/-6$), with lower limits to the inner disc radius of 20--30 \rg\ and $i<17$ deg; the other parameters are consistent within the errors with the \xillver\ fit.

\subsection{The broad-band fit I. Testing relativistic reflection}\label{subsec:refl_fits}
\begin{figure} 
        \includegraphics[width=\columnwidth]{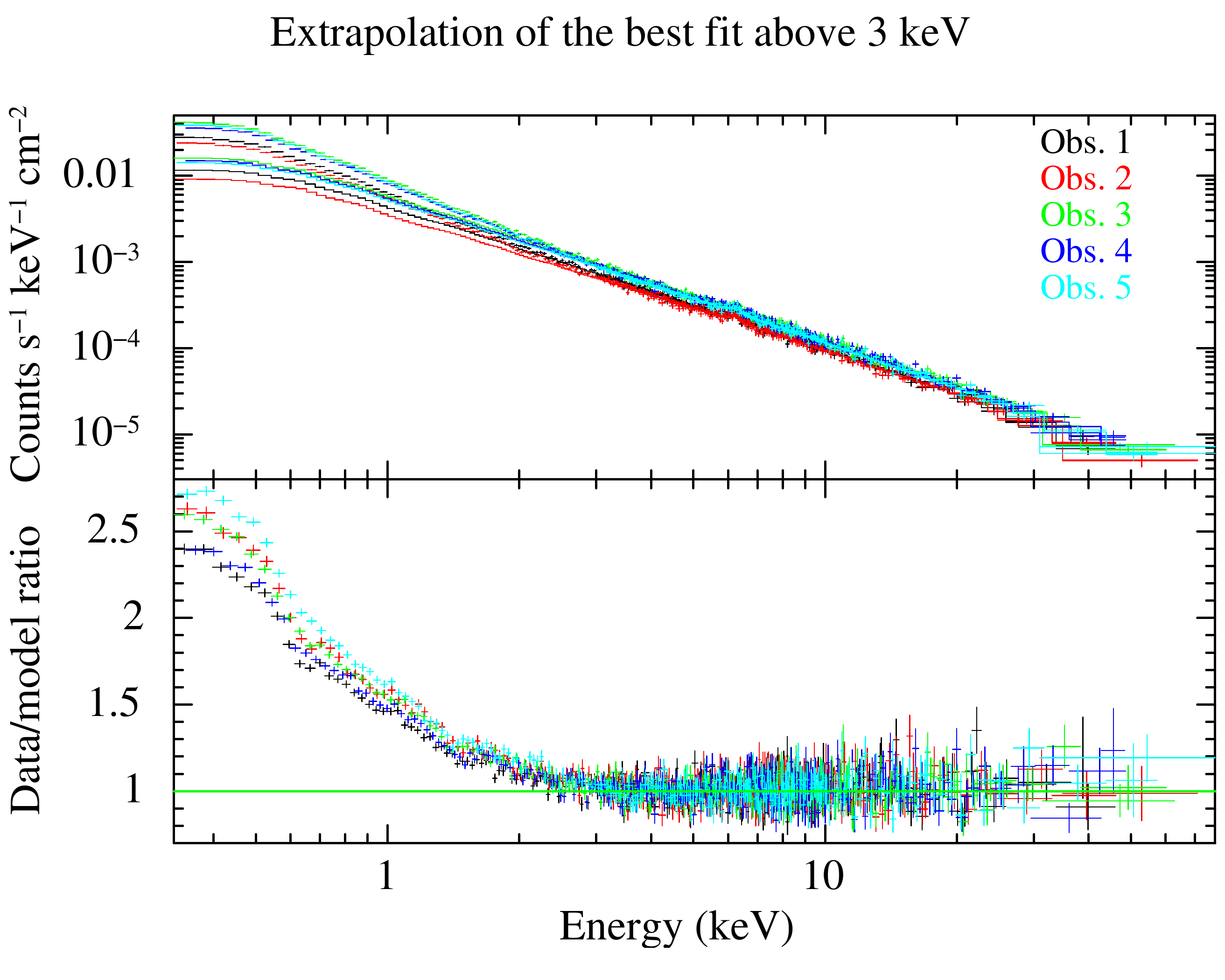}
        \caption{\label{fig:extrap} Upper panel: pn and \nus~spectra fitted with a cut-off power law plus \xillver\ in the 3--79 keV band (see Sect. \ref{subsec:refl}). Lower panel: Ratio of the broad-band spectra, down to 0.3 keV, to the model. The data were binned for plotting purposes.}
\end{figure}

After obtaining constraints on the reflection component, we proceeded to fit the pn and \nus\ data in the whole X-ray energy band (0.3--79 keV). 
Extrapolating the best-fitting model (cut-off power law plus \xillver) above 3 keV to lower energies, we found a significant soft excess  (Fig. \ref{fig:extrap}).
In addition,  refitting the data in the 0.3--79 keV band and including a 0.5 keV Gaussian emission line (see Appendix \ref{app:rgs}), we found that the fit using \xillver\ or \relxill\ only is very poor ($\rchisq>2.5$), with significant residuals below 1 keV. 

We then tested a broad-band model including the primary continuum plus two reflection components. We used \relxill\ to model the continuum plus the ionized reflection from the inner disc, which produces a soft excess. For the constant reflection component we tested two different models, namely \xillver\ and \borus, as we describe below.
We note that when using these models an additional 0.5 keV Gaussian line is not required. 

\paragraph{Model A: \relxill+\xillver.}
In \relxill, the emissivity index $q$, the inner disc radius, and the ionization were left free, but were tied among the observations, with no significant improvement by leaving them free to change. The photon index of the continuum, the cut-off energy, the reflection fraction, and the normalization were all free to vary among the observations. The outer disc radius was fixed at 400 \rg. In \xillver, the ionization and the normalization were free and tied among the observations; the photon index and the cut-off energy were instead fixed at the average values of the continuum in \relxill. We also left free (but tied among the observations) the iron abundance and the inclination, assuming the same values for \relxill\ and \xillver.

\paragraph{Model B: \relxilld+\xillverd.}
We tested the flavour of \relxill\ and \xillver\ that provides the density of the reflecting material as a free parameter \citep{xillverD}. We left  the disc density in both \relxilld\
and \xillverd\ free but tied among the observations, without imposing a link between the two components. The other parameters were set as in model A, with the difference that in the current version of \relxilld\ and \xillverd\ the cut-off energy is fixed at 300 keV.

\paragraph{Model C: \relxill+\borus.}
We replaced \xillver\ with \borus\ \citep{borus}, which describes neutral reflection from a gas torus. This model includes Compton scattering plus self-consistent fluorescent line emission. 
In \borus, the half-opening angle of the torus $\theta\subrm{tor}$ and the normalization were free and tied among the observations, while the photon index and cut-off energy were fixed at the \relxill\ average values. The iron abundance and the inclination were linked between \relxill\ and \borus, and tied among the observations. We note that the inclination angle is defined in the same way in \relxill\ and \borus, being measured from the symmetry axis in both cases. The other parameters of \relxill\ were set as in model A. 

\begin{figure}
        \includegraphics[width=\columnwidth]{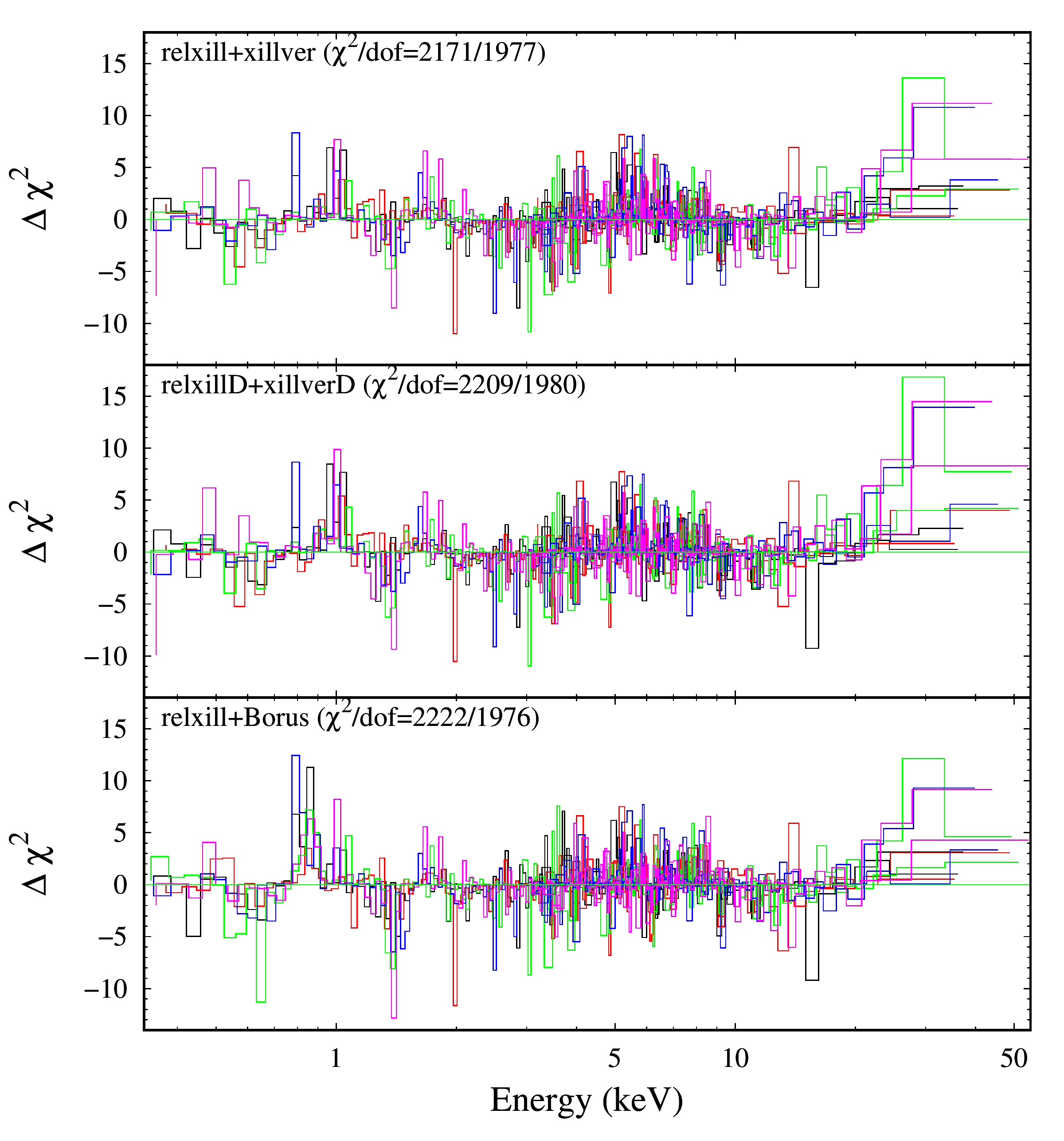}
        \caption{
                 Contributions to $\chisq$ for the relativistic reflection fits (rebinned for plotting purposes) discussed in Sect. \ref{subsec:refl_fits}.
                }
        \label{fig:chi-refl}
\end{figure}

\paragraph{}
We list the best-fitting parameters for each model in Table \ref{tab:params_refl}. In Fig. \ref{fig:chi-refl} we show the residuals for the three fits. 

Model A provides the best fit in a statistical sense, having $=2171/1977$  $(1.098)$. 
However, the model leaves significant positive residuals in the high-energy band above 20 keV. The best-fitting photon indices of the continuum are relatively steep, being 2.0--2.1. Related to this, the cut-off energy is mostly found to have lower limits $>500$ keV. Very similar results are obtained replacing \relxill\ and \xillver\ with the corresponding Comptonization flavours, \relxillcp\ and \xillvercp, thus fitting for the electron temperature instead of the exponential cut-off. \relxillcp+\xillvercp\ actually yield a worse fit with $\dchi=+50$ for the same number of dof, and mostly lower limits to the temperature.
Model B yields $\rchisq=2209/1980$  $(1.116)$ with significant, positive residuals above 20 keV. The fixed cut-off energy at 300 keV likely explains the worse fit compared with model A, as well as the tight constraints on the photon index (which is $\sim 2.0$, as in model A) and on the reflection fraction. The pn--\nus\ discrepancy  in photon index is essentially zero, which is unexpected (see Appendix \ref{app:pn-nustar}).

Finally, model C yields $\rchisq=2222/1976$  $(1.124)$, with positive residuals above 20 keV and at 0.8--0.9 keV. The latter are possibly due to the lack of soft emission lines in the neutral reflection component \borus\ (while they are included in \xillver). This model also requires some extreme parameters. In \relxill, the inner radius is pegged at the minimum value of 1.235 \rg\ with an upper limit of 1.3 \rg, and the emissivity index is pegged at 10 with a lower limit of 9. In \borus, the half-opening angle has a lower limit of 76 deg and the column density is pegged at the maximum value $\log \nh = 25.5$. Also, the iron abundance is $<0.52$ and the inclination is tightly constrained at 67--68 deg. 

We conclude that model A (\relxill+\xillver) provides the most satisfactory fit to the data, even if only assuming that the inclination is very low (pegged at the minimum value of 3 deg, with an upper limit of only 5 deg; fixing
the inclination at 30 deg, we obtained a worse fit with
$\Delta \chisq = +25$ for 1 additional dof), and the spin very high
(given the small inner disc radius, \aer{1.64}{0.06}{0.04} \rg).


\begin{table*} 
        \centering
        \caption{  Best-fitting parameters of the three models described in Sect. \ref{subsec:refl_fits}. In the second column we list the fit parameters that were tied among all observations. In the subsequent columns, we give the fit parameters that were free to vary for each observation. 
                \label{tab:params_refl}
        }
        \begin{tabular}{l c c c c c c}
                \hline & all obs. & obs. 1 & obs. 2 & obs. 3 & obs. 4 & obs. 5 \\
                \hline
        
        \multicolumn{7}{c}{model A: \relxill+\xillver}\\
        $i$ (deg)& $<5^{\dagger}$&&&&&\\ 
        $A\subrm{Fe}$ (solar)&\ser{0.8}{0.2}&&&&&\\
        $\Gamma$&&\aer{2.03}{0.02}{0.03}&\ser{2.01}{0.03}&\ser{2.13}{0.02}&\ser{2.09}{0.02}&\aer{2.13}{0.03}{0.02}\\
        $\cut$ (keV)&&\aer{230}{170}{80}&$>500$&$>700$&$>600$&$>800$\\
        $\refl$&&\aer{0.8}{0.2}{0.1}&\aer{0.73}{0.06}{0.08}&\aer{0.9}{0.2}{0.1}&\aer{0.8}{0.2}{0.1}&\aer{1.0}{0.2}{0.1}\\ 
        $\log \xi_{\textsc{relxill}}$ (\lumcgs\ cm)&\aer{2.4}{0.3}{0.1}&&&&&\\  
        $q$&\ser{5.8}{0.2}&&&&&\\
                $R\subrm{in}$ (\rg)&\aer{1.64}{0.06}{0.04}&&&&&\\                            
        $N_{\textsc{relxill}}$&&\aer{8.8}{0.7}{0.6}&\aer{9.1}{0.4}{1.5}&\ser{13.0}{0.5}&\ser{12.5}{0.5}&\aer{11.9}{0.3}{0.8}\\    
        \noalign{\medskip}
        $\log \xi_{\textsc{xillver}}$ (\lumcgs\ cm)&\aer{1.22}{0.09}{0.07}&&&&&\\  
        $N_{\textsc{xillver}}$&\aer{6.8}{0.3}{0.4}&&&&&\\ 
        \noalign{\medskip}    
        $\Delta \Gamma$&\ser{0.04}{0.02}&&&&&\\           
        \noalign{\medskip}              
        $\rchisq$&2171/1977&&&&&\\  
                \hline  
        
        \multicolumn{7}{c}{model B: \relxilld+\xillverd}\\
        $i$ (deg)&$<5^{\dagger}$&&&&&\\ 
        $A\subrm{Fe}$ (solar)&\ser{0.8}{0.1}&&&&&\\
        $\Gamma$&&\ser{2.005}{0.005}&\ser{1.965}{0.005}&\ser{2.078}{0.005}&\ser{2.038}{0.005}&\ser{2.081}{0.005}\\
        $\refl$&&\ser{0.60}{0.04}&\ser{0.58}{0.04}&\ser{0.72}{0.03}&\ser{0.66}{0.03}&\ser{0.77}{0.03}\\ 
        $\log \xi_{\textsc{relxillD}}$ (\lumcgs\ cm)&\ser{2.30}{0.02}&&&&&\\  
        $q$&\ser{5.7}{0.1}&&&&&\\
        $\log n_{\textsc{relxillD}}$ (cm$^{-3}$)&\aer{17.03}{0.01}{0.21}&&&&&\\  
        $R\subrm{in}$ (\rg)&\ser{1.68}{0.04}&&&&&\\                            
        $N_{\textsc{relxillD}}$ (\tento{-5})&&\ser{9.5}{0.1}&\ser{8.7}{0.1}&\ser{12.4}{0.1}&\ser{11.8}{0.1}&\ser{11.4}{0.1}\\    
        \noalign{\medskip}
        $\log \xi_{\textsc{xillverD}}$ (\lumcgs\ cm)&\ser{1.23}{0.07}&&&&&\\ 
        $\log n_{\textsc{xillverD}}$ (cm$^{-3}$)&\aer{16.98}{0.03}{0.08}&&&&&\\   
        $N_{\textsc{xillverD}}$ (\tento{-5})&\aer{5.9}{0.2}{0.8}&&&&&\\          
        \noalign{\medskip}         
        $\Delta \Gamma$&\ser{0.003}{0.003}$^{\dagger}$&&&&&\\           
        \noalign{\medskip}         
        $\rchisq$&2209/1980&&&&&\\              
                \hline
        
        \multicolumn{7}{c}{model C: \relxill+ \borus}\\ 
        $i$&\aer{67.5}{1.3}{1.9}&&&&&\\ 
        $A\subrm{Fe}$ (solar)&$<0.52$&&&&&\\
        $\Gamma$&&\ser{2.15}{0.02}&\ser{2.13}{0.02}&\ser{2.22}{0.01}&\aer{2.18}{0.02}{0.01}&\ser{2.22}{0.01}\\
        $\cut$ (keV)&&\aer{210}{280}{70}&$>200$&$>750$&$>500$&$>700$\\
        $\refl$&&\aer{0.69}{0.05}{0.08}&\aer{0.68}{0.04}{0.07}&\aer{0.68}{0.18}{0.07}&\aer{0.66}{0.03}{0.07}&\aer{0.74}{0.04}{0.07}\\ 
        $\log \xi_{\textsc{relxill}}$ (\lumcgs\ cm)&\aer{1.3}{0.1}{0.3}&&&&&\\  
        $q$&$>9$&&&&&\\
        $R\subrm{in}$ (\rg)&$<1.3$&&&&&\\                            
        $N_{\textsc{relxill}}$&&\aer{10.4}{0.6}{0.7}&\aer{9.9}{0.4}{1.1}&\aer{15.6}{0.3}{0.2}&\aer{14.4}{0.3}{0.5}&\aer{14.4}{0.3}{0.2}\\    
        \noalign{\medskip}
        $\theta\subrm{tor}$ (deg)&$>76$&&&&&\\ 
        $\log N\subrm{H,tor}$ (\sqcm)&$>25.4^{\dagger}$&&&&&\\   
        $N_{\textsc{borus}}$&\ser{0.04}{0.02}&&&&&\\  
        \noalign{\medskip} 
        $\Delta \Gamma$&\aer{0.08}{0.02}{0.01}&&&&&\\           
        \noalign{\medskip}         
        $\rchisq$&2222/1976&&&&&\\      
        \hline                  
        \end{tabular}
        \tablefoot{
                \tablefoottext{$\dagger$}{Constraint estimated from the 1D contour plot (\textsc{steppar} command in \xspec).}
        }
\end{table*}

\subsection{The broad-band fit II. Testing the two-corona scenario}\label{subsec:fit}
In the next step we tested a two-corona scenario, in which warm Comptonization accounts for the soft excess. We fitted the pn and \nus\ data in the 0.3--79 keV band, also including the optical--UV data from the OM.  The different components of the model are described below. 
\paragraph{The primary continuum and soft excess:}
The hard X-ray spectrum is modelled with the thermal Comptonization model \nthcomp. We left  the electron temperature $\kte$ and the photon index $\Gamma$ of the asymptotic power law free to vary among the different observations. For the seed photons, we assumed a multicolour disc black-body distribution \cite[][]{mitsuda1984,makishima1986} and left the seed temperature $\ktbb$ free, but tied among the observations.
We used \nthcomp\ also to model the soft excess, fitting for the electron temperature and the photon index. The model thus included a hot \nthcomp\ component with electron temperature $\kteh$ and photon index $\gammah$, and a warm \nthcomp\ component with temperature $\ktew$ and photon index $\gammaw$ (see Table \ref{params}). The seed temperature $\ktbb$ was the same for both components. 
\paragraph{Reflection:}
Following Sect. \ref{subsec:refl}, to model the reflection component we used \xillvercp\, namely the flavour of \xillver\ in which the illumination spectrum is modelled with \nthcomp\ instead of a cut-off power law.
The free parameters of \xillvercp\ were the iron abundance, the ionization parameter, and the normalization. These parameters were kept tied to a common value as they were consistent with being constant. We fixed the photon index and electron temperature of the incident spectrum to the average values found in each observation for the primary continuum (see also Sect. \ref{subsec:refl}).
\paragraph{Soft emission lines:}
We included a narrow Gaussian line to account for positive residuals around 0.5 keV that can be associated with the \ovii\ complex detected in the RGS spectra (Appendix \ref{app:rgs}). 
\paragraph{Small blue bump:} 
This broad feature is generally observed in the optical--UV spectrum of AGNs, in the 2000--4000 \AA\ band, and is due to a blend of strong \feii\ lines and the Balmer continuum emission \cite[][]{grandi1982,wills1985}. To account for this component, we produced a table model for \xspec\ (\smallbb) using the calculations of \cite{wills1985} and \cite{grandi1982} for the \feii\ lines and for the Balmer continuum, respectively. 
The model flux of this component was found to be \serexp{8.3}{0.7}{-12} \fluxcgs, and is consistent with being constant among the different observations.

\begin{figure} 
        \includegraphics[width=\columnwidth]{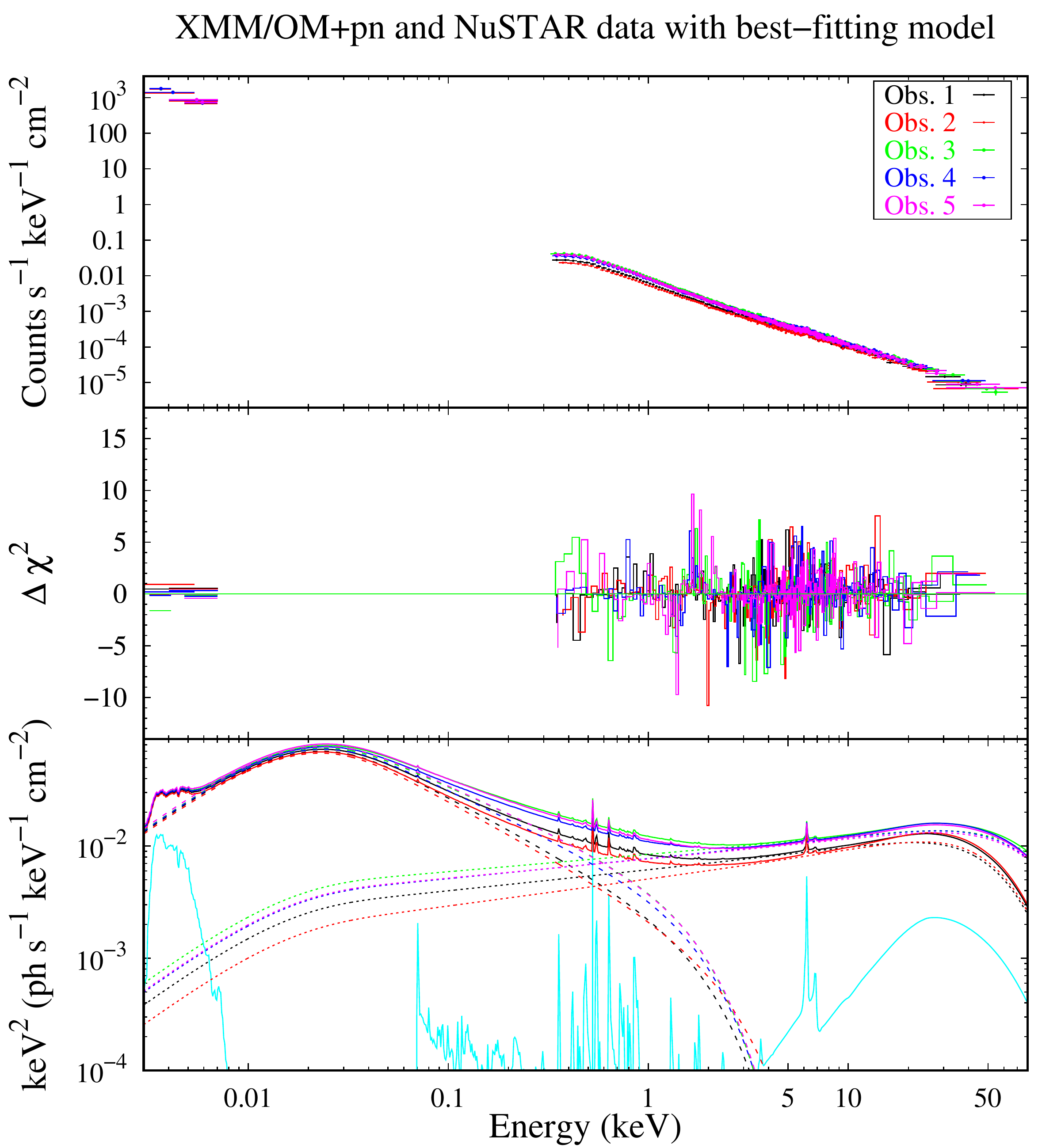}
        \caption{\label{fig:fit} Broad-band UV--X-ray data and best-fitting model discussed in Sect. \ref{subsec:fit} (see Table \ref{params}). Upper panel: \xmm/OM, pn, and \nus\ data (rebinned for plotting purposes) with folded model. 
        Middle panel: Contribution to $\chi^2$. Bottom panel: Best-fitting model $E^2 f(E)$, without absorption, with the plot of the warm and hot \nthcomp\ components (dashed and dotted lines, respectively), the small blue bump, the Gaussian line at 0.54 keV, and the reflection component (cyan solid lines).
                }
\end{figure}

\begin{table*}
\centering
                \caption{Best-fitting parameters of the broad-band model described in Sect. \ref{subsec:fit}: {\sc smallBB+zgauss+nthcomp,w+nthcomp,h+xillverCp} in \xspec\ notation. In the second column we list the fit parameters that were tied among all observations. In the subsequent columns, we list  the fit parameters that were free to vary for each observation. The \zga\ normalization is the line flux density in units of \fluxph. The \nthcomp\ normalization is the flux density in units of \fluxph\ at 1 keV. For the definition of the \xillvercp\ normalization, see \cite{dauser2016}.
                        \label{params}
                }
                \begin{tabular}{l c c c c c c}
                        \hline & all obs. & obs. 1 & obs. 2 & obs. 3 & obs. 4 & obs. 5 \\
                        \hline
                        $F_{\textsc{smallBB}}$ ($10^{-12}$ \fluxcgs)& \ser{8.3}{0.7} &&&&& \\ 
                        %\hline 
                        \noalign{\medskip} 
                        $E_{\textsc{zgauss}}$ (keV) &\ser{0.543}{0.008}&&&&& \\ 
                        $N_{\textsc{zgauss}}$ ($10^{-4}$) &\ser{2.5}{0.4}&&&&& \\ 
                        \noalign{\medskip}
                        $\gammaw$ && \ser{ 2.91 }{ 0.04 } & \ser{ 2.94 }{ 0.05 }  & \ser{ 2.76 }{ 0.04 } & \ser{ 2.81 }{ 0.04 } & \ser{ 2.77 }{ 0.04} \\
                        $\ktew$ (keV) && \aer{0.45}{0.09}{ 0.06 } & \aer{ 0.6 }{ 0.3 }{0.1}  & \aer{ 0.40 }{ 0.06 }{ 0.05 }  & \aer{0.42}{0.06}{0.05} & \aer{ 0.41 }{ 0.04 }{0.05}\\
                        $N_{\textsc{nthcomp,w}}$ ($10^{-3}$)&& \ser{2.2}{0.3} 
                        & \ser{ 2.1 }{ 0.3 }  & \ser{ 3.7 }{ 0.4 }  & \ser{ 3.2 }{ 0.4 } & \ser{ 3.7 }{ 0.4 }\\
                        $\ktbb$ (eV) & \ser{7.0}{0.5} &&&&& \\
                        \noalign{\medskip} 
                        $\gammah$ && \ser{ 1.82 }{ 0.02 }& \ser{ 1.76 }{0.02}  & \ser{ 1.85}{ 0.02 } & \ser{ 1.82}{ 0.02 } & \ser{ 1.83}{ 0.02 } \\
                        $\kteh$ (keV) && \aer{ 13 }{ 7}{ 3 } & \aer{ 13 }{ 6 }{ 3 }  & \aer{ 25 }{ 75 }{ 8 }  & \aer{ 20 }{ 80 }{ 6 }  & \aer{ 20 }{70 }{ 6 } \\
                        $N_{\textsc{nthcomp,h}}$ ($10^{-3}$)&& \ser{6.2}{0.3} & \ser{5.6}{0.3} & \ser{ 7.9 }{ 0.3 }  &  \ser{ 7.8 }{ 0.3 }    & \ser{ 7.6 }{ 0.3 }  \\
                        \noalign{\medskip} 
                        $N_{\textsc{xillverCp}}$ (\tento{-5}) & \aer{ 2.9}{ 0.4}{0.5} &&&&& \\
                        $\afe$ & \aer{ 2.8 }{ 0.7 }{ 0.6 } &&&&& \\
                        $\log \xi$ (\lumcgs\ cm) & \aer{ 1.71 }{ 0.05 }{ 0.17 } &&&&&\\
                        \noalign{\medskip} 
                        $\ftwoten$ (\tento{-11} \fluxcgs) &&\ser{2.23}{0.03}&\ser{2.11}{0.03}& \ser{2.80}{0.04}&\ser{2.76}{0.04}&\ser{2.69}{0.04} \\
                        \noalign{\medskip} 
                        $\lbol$ (\tento{44} \lumcgs) &&\ser{8.63}{0.03}&\ser{8.16}{0.03}&\ser{10.36}{0.03}&\ser{9.98}{0.03}&\ser{10.49}{0.03} \\
                        \noalign{\medskip}
                        $\Delta \Gamma$ &\ser{0.07}{0.02}&&&&&\\
                        \noalign{\medskip}
                        $\rchisq$ &$2048/1986$&&&&&\\
                        \hline                  
                \end{tabular}
\end{table*}

We show the data, residuals, and best-fitting model in Fig. \ref{fig:fit}, while all the best-fitting parameters are listed in Table \ref{params}. We obtain $\rchisq = 2048/1986$. Considering the X-ray data only, this corresponds to $\rchisq=2042/1968$ ($\dchi/\ddof=-129/-9$ compared with the \relxill+\xillver\ fit). The pn--\nus\ discrepancy in photon index is $\Delta \Gamma = 0.07$.

\begin{figure} 
        \includegraphics[width=\columnwidth]{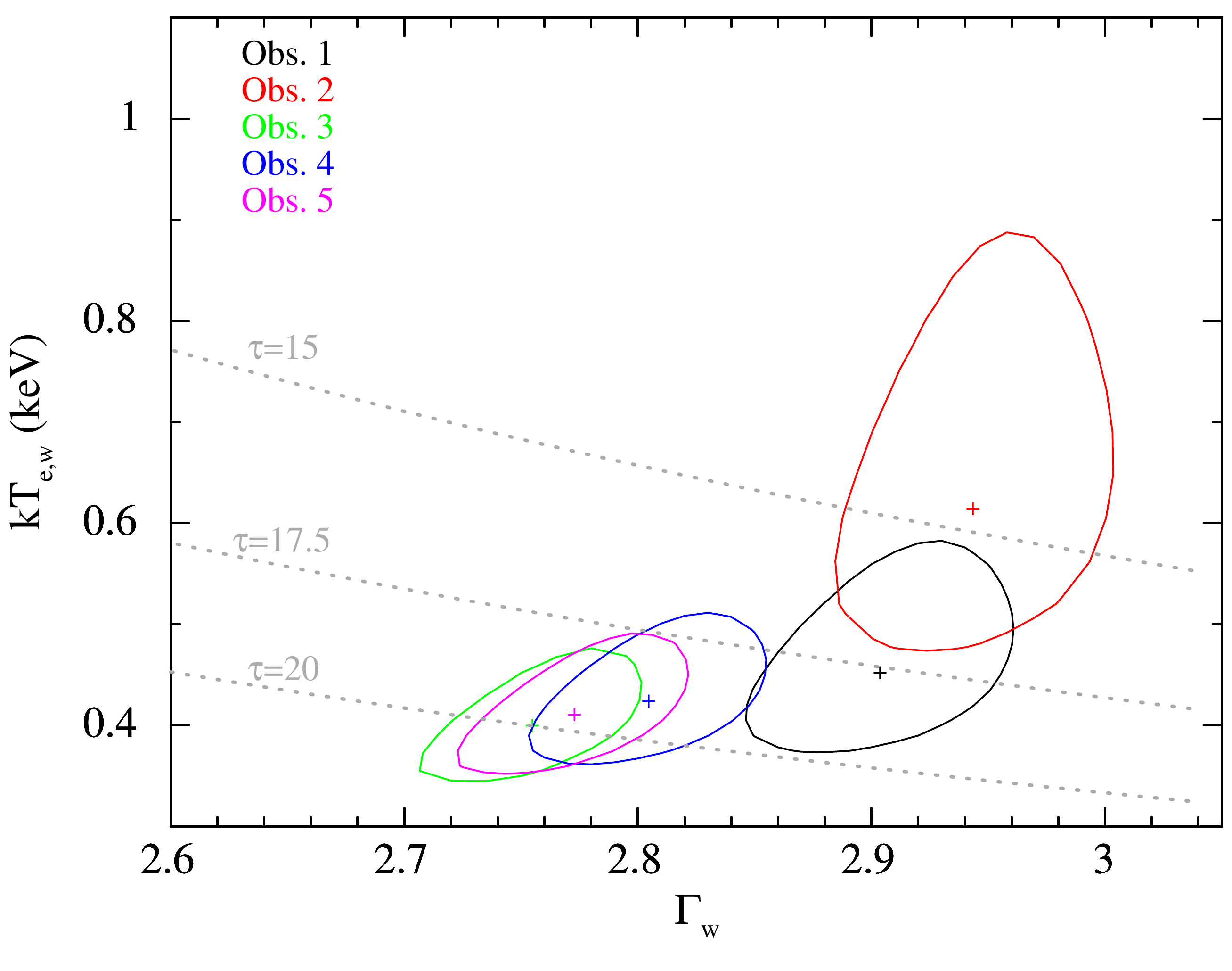}
        \caption{\label{fig:cont_warm} Contour plots of the electron temperature vs.  photon index of the warm corona at the 90\% confidence level. 
                Grey dotted lines correspond to contours of constant optical depth.}
\end{figure}

\begin{figure} 
        \includegraphics[width=\columnwidth]{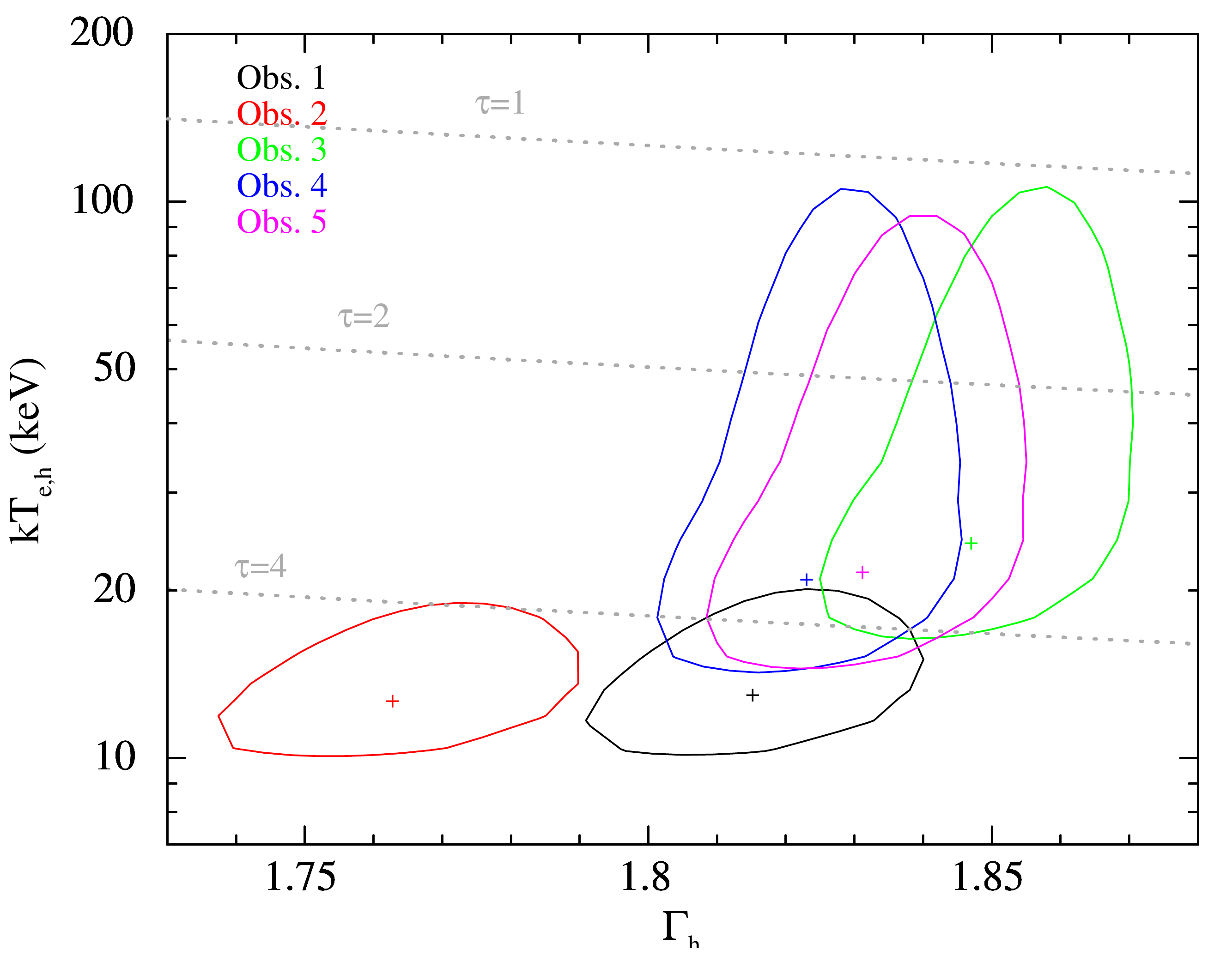}
        \caption{\label{fig:cont_hot} Contour plots of the electron temperature vs. photon index of the hot corona at the 90\% confidence level. 
                Grey dotted lines correspond to contours of constant optical depth.}
\end{figure}

For the warm \nthcomp\ component, we find a variable photon index in the range 2.7--3.0 and a temperature in the range 0.4--0.8 keV (see Fig. \ref{fig:cont_warm}). The corresponding optical depth, as derived from the \nthcomp\ model\footnote{
                The optical depth $\tau$ in \nthcomp\ is related to the photon index $\Gamma$ of the asymptotic power law and to the temperature $\Theta \equiv \kte/m\subrm{e}c^2$ via the formula $\tau = \{ 2.25 + 3/[\Theta \times (\Gamma+0.5)^2 - 2.25]\}^{0.5} - 1.5$.
        }, 
is roughly consistent with 17.5 with some hints of variability between observations 2 and 3. 
The hot \nthcomp\ component is nearly constant in spectral shape, with the exception of observation 2, which has a flatter photon index (see Fig. \ref{fig:cont_hot}). 
Given the uncertainties, the temperature is roughly consistent with being always $\sim 20$ keV,
while the optical depth is consistent with $\sim 4$.  
However, the flatter photon index in observation 2 suggests a slightly greater optical depth (and possibly a smaller temperature).

\begin{figure} 
        \includegraphics[width=\columnwidth]{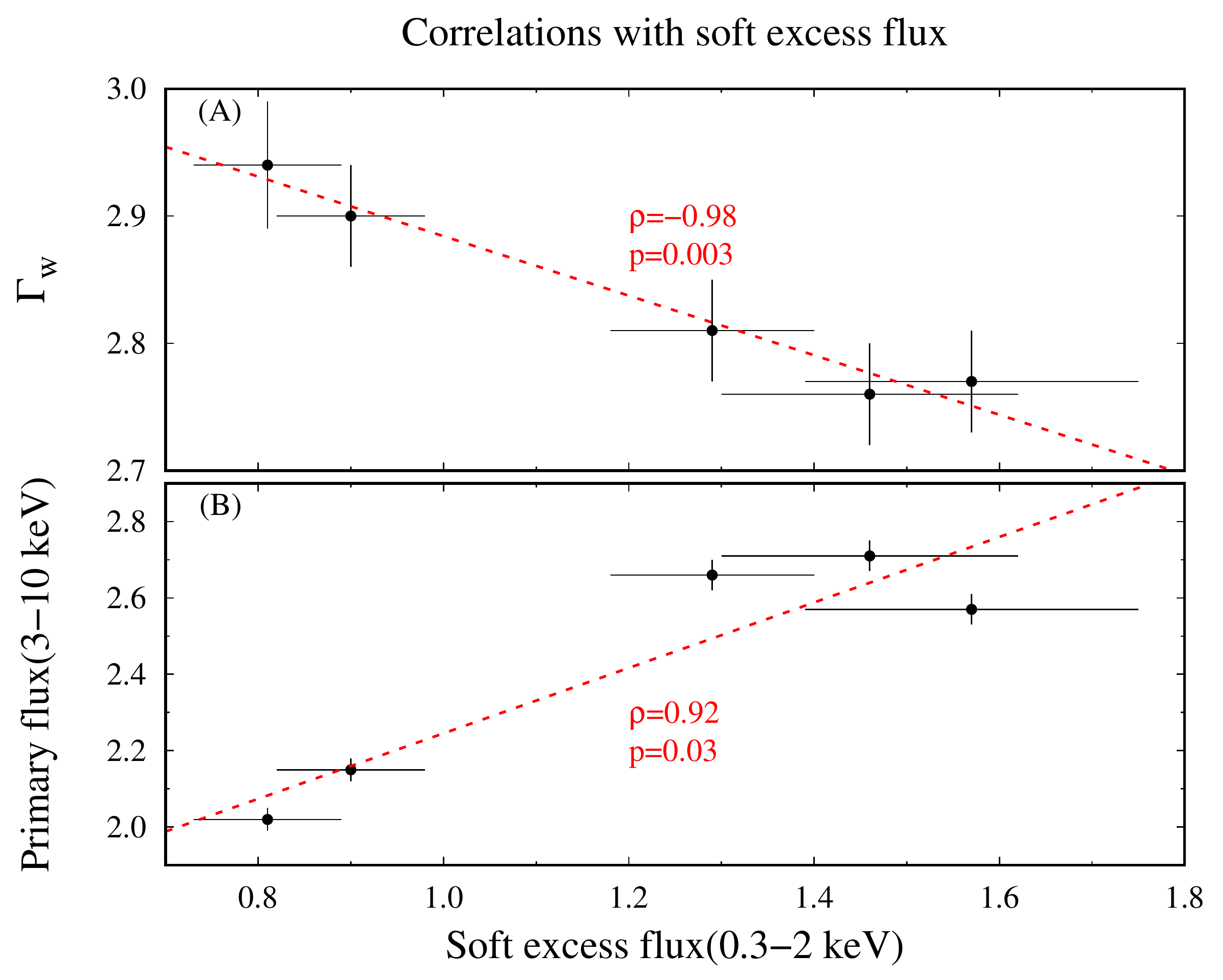}
        \caption{\label{fig:fluxes}
                Panel (A): Photon index $\gammaw$ vs. flux (0.3--2 keV) of the warm \nthcomp\ component, for the different observations.
                 Panel (B): Primary X-ray flux (3--10 keV) of the hot \nthcomp\ component vs. flux of the warm \nthcomp\ component. The fluxes are in units of \tento{-11} \fluxcgs. Red dashed lines represent linear fits to the data.}
\end{figure}

The warm \nthcomp\ photon index is significantly anticorrelated with the flux (see Fig. \ref{fig:fluxes}, panel A): the Pearson's correlation coefficient is $-0.98$ with a $p$-value of 0.003. Also, the flux of the hot \nthcomp\ component in the 3--10 keV band is correlated with the flux of the warm \nthcomp\ component in the 0.3--2 keV band (see Fig. \ref{fig:fluxes}, panel B), with a Pearson's correlation coefficient of 0.92 and a $p$-value of 0.03. 
This indicates a correlation between the primary X-ray emission from the hot corona and the soft excess.

The absorption-corrected model luminosities in the 0.001-1000 keV band are in the range $0.8{-}1.0 \times 10^{45}$ \lumcgs.
Assuming a black hole mass of \expo{4}{7} \msun, we obtain an Eddington ratio $ L/L_{\textrm{Edd}} \simeq 0.16-0.20$.
The seed photon temperature of both \nthcomp\ components is found to be around 7 eV. This temperature is expected at a radius of $\sim 10$ \rg\ in a standard \cite{ss1973} accretion disc, for the black hole mass and accretion rate above. 

To further investigate the nature of the accretion flow, we tested a physically motivated model for the hot corona: the jet-emitting disc \citep[JED:][]{jed1,jed2,jed3}, originally proposed for X-ray binaries. 
Assuming scale invariance for accreting black hole systems, the JED can be used to model the high-energy emission of AGNs by simply changing the black hole mass.

\subsection{Testing the jet-emitting disc}\label{subsec:jed}
The JED model has been developed to explain the different spectral states observed in X-ray binaries \citep{jed1}. In particular, the model is able to explain the X-ray spectral properties and the presence of radio jets observed during hard states \citep{pop2010}. Although \he\ is classified as a radio-quiet Seyfert, it shows an unresolved radio emission with a flux density of \ser{9.5}{0.6} mJy at 1.4 GHz \citep[from \textit{Very Large Array} data:][]{condon1998} which is consistent with the prediction of the so-called fundamental plane of black hole activity \citep{merloni2003}, given the observed X-ray luminosity and black hole mass of the source.

The JED paradigm assumes the existence of a large-scale vertical magnetic field 
$B_z$ that can become dominant in the innermost region of the disc, allowing the production of self-confined jets. 
The capability of generating jets strongly depends on the disc magnetization, defined as the ratio $B_z^2/\mu_0P_{tot}$, where $P_{tot}$ is the gas plus radiation pressure. Stationary jets are only produced at high disc magnetization, of order unity \citep{ferreira1995,pop2008}. A much lower magnetization would thus correspond to a standard accretion disc (SAD). Within this paradigm, the accretion flow has two constituents: an outer SAD, extending down to a transition radius $r_J$ where the magnetization becomes of order unity, and an inner JED down to the innermost stable circular orbit (ISCO).
When the transition radius is close to the ISCO, the thermal emission from the SAD dominates and the source is in a soft state; for large transition radii, the X-ray emission from the JED dominates and the source is in a hard state (for details, see \citealt{jed3}). In addition to  the transition radius, the other crucial parameter is the accretion rate, which determines the total luminosity and influences the aspect ratio of the disc height to radius \citep{jed2}.
The evolution of transition radius and accretion rate can explain the spectral properties of X-ray binaries like GX 339-4 during outbursting cycles \citep{jed3,jed4}. 

We tested the SAD--JED model, for the first time on an AGN, following \cite{jed2,jed3}, who developed a two-temperature plasma code which includes advection and radiation losses via Bremsstrahlung, synchrotron, and Comptonization.  These radiative processes are taken into account through the \textsc{belm} code \citep{belm1,belm2}.
The effect of the illumination on the inner JED by cold photons from the outer SAD is also taken into account \citep{jed3}. 
We fitted the \xmm\ and \nus\ data as in Sect. \ref{subsec:fit}. 
We produced two \xspec\ tables, one for the JED and one for the SAD, which both have the same two parameters $r_J$ and $\mdot$. Having two separate tables allows us to model the soft excess as a Comptonized tail of the SAD alone.
The different components of the model are described below. 
\paragraph{The primary continuum:} In the SAD--JED scenario translated to AGNs, the SAD produces the optical--UV bump while the JED produces the bulk of the X-ray emission. The free parameters of these models are the SAD--JED transition radius $r_J$ and the accretion rate $\mdot$ at the ISCO in Eddington units ($\dot{M}_{\textrm{Edd}} \equiv L_{\textrm{Edd}}/c^2$, thus $\mdot \equiv \dot{M}/\dot{M}_{\textrm{Edd}}  = L/\eta L_{\textrm{Edd}}$, where $\eta$ is the efficiency of mass-energy conversion.)\footnote{The JED accretion rate is linked to the SAD accretion rate via $ \mdot (r)/ \mdot_{\textrm{SAD}} =(r/r_J)^{\xi}$, where $\xi$ is the ejection efficiency (fixed at 0.01).}.
No link is imposed a priori between $r_J$ and $\mdot$ in the SAD--JED model.
We left $r_J$ and $\mdot$, which determine the broad-band spectral shape, free to change among the different observations.
The observed flux is not a free parameter per se, because it depends only on the total luminosity (which is mostly set by the accretion rate and the black hole mass) and on the distance of the object. We assumed a distance of 144 Mpc, as obtained from the redshift and assuming a standard cosmology with $H_0 = 70$ km \pers, $\Omega_M=0.3$, and $\Omega_{\Lambda}=0.7$. To allow for flux normalization, we left also the black hole mass $\mbh$ free to vary (but tied among the different observations). 
The SAD--JED model also has  several dynamical parameters related to the production of jets, and thus not relevant for the present work.
We thus used the same parameters that were chosen by \cite{jed4} to reproduce an outburst from the X-ray binary GX 339-4: the disc magnetization, set to 0.5, which does not strongly influence the X-ray emission \citep{jed2}; the local JED ejection efficiency, set to 0.01; the sonic Mach number, set to 1.5; the fraction of the accretion power channelled into the jet, set to 0.3.  
The definition of these parameters and a study of their effects can be found in \cite{jed2,jed3}.

\paragraph{The soft excess:} Motivated by the results of the broad-band spectral analysis given in Sect. \ref{subsec:fit}, we modelled the soft excess as a Comptonized tail of the SAD emission. We thus convolved the SAD component with the model \simpl\ in \xspec\ \citep{simpl}, in which a fraction of the input photons is scattered into a power law component. We left the photon index of \simpl\ free to vary among the observations, in close analogy to the \nthcomp\ fit discussed above. In \simpl, we assumed a scattered fraction of 1, with no improvement found when leaving this parameter free.
Since the model \simpl\ currently available in \xspec\ does not take into account the roll-over due to the finite temperature of the scattering medium, we also included an exponential cut-off. This is an acceptable approximation, since our main goal was to get insights into the nature of the hot corona. 
\paragraph{Reflection:} Again motivated by the results of Sect. \ref{subsec:fit}, we included a reflection component modelled with \xillvercp. 
We fixed the parameters at the best-fitting values found in Sect. \ref{subsec:fit}, with no significant improvement by leaving them free to vary.

\paragraph{Soft emission line and small blue bump:} 
We included these components, as in Sect. \ref{subsec:fit}. We note that the small blue bump is related to the broad-line region and is not reproduced by the SAD component.

\paragraph{} We found a good fit with $\rchisq=2052/1999$ (i.e. $\dchi/\ddof=+4/+13$ compared with the two-corona fit). The electron temperature and Thomson optical depth of the JED are computed as a function of radius assuming thermal equilibrium \citep{jed3}. We obtained a temperature of roughly $\sim \tentom{8}$ K and an optical depth of 3-4, i.e. of the same order of magnitude as the values found in Sect. \ref{subsec:fit}. The physical properties of the SAD--JED solution are discussed in detail in Appendix \ref{app:jed}. In Fig. \ref{fig:jed} we plot the best-fitting SAD and JED components. 
We note that this SAD--JED model has only five free parameters. One, the black hole mass, is constant among the observations, while four are variable:  $r_J$, $\mdot$, the soft excess photon index $\Gamma_s$, and its cut-off energy $E_s$ (we use the subscript $s$ to distinguish these parameters from those of the warm \nthcomp\ used in Sect. \ref{subsec:fit}). 
In the SAD--JED  model there is no need for a free normalization because the model luminosity is a function of the black hole mass and the other free parameters. Assuming a luminosity distance of 144 Mpc, the observed flux is 
correctly reproduced with a best-fitting black hole mass of \serexp{3.69}{0.07}{7} \msun.
This value of the mass is consistent with that estimated from the X-ray variability and from the H $\beta$ line (Sect. \ref{sec:timing}). 
The SAD--JED transition radius is between 18 and 20 gravitational radii, while the accretion rate varies between 0.7 and 0.9 in units of $L\subrm{Edd}/c^2$. No significant correlation is found between these two parameters, although there is a hint of a lower accretion rates for smaller transition radii (see Fig. \ref{fig:pjed}). 
A correlation between the accretion rate and the transition radius has been found during the intermediate state of GX 339-4; this is briefly 
discussed by \cite{jed4}  (see their Fig. 7).
However, in our case the error bars are of the same order of magnitude as the variations, thus preventing us from drawing any strong conclusions. 

\begin{figure} 
        \includegraphics[width=\columnwidth]{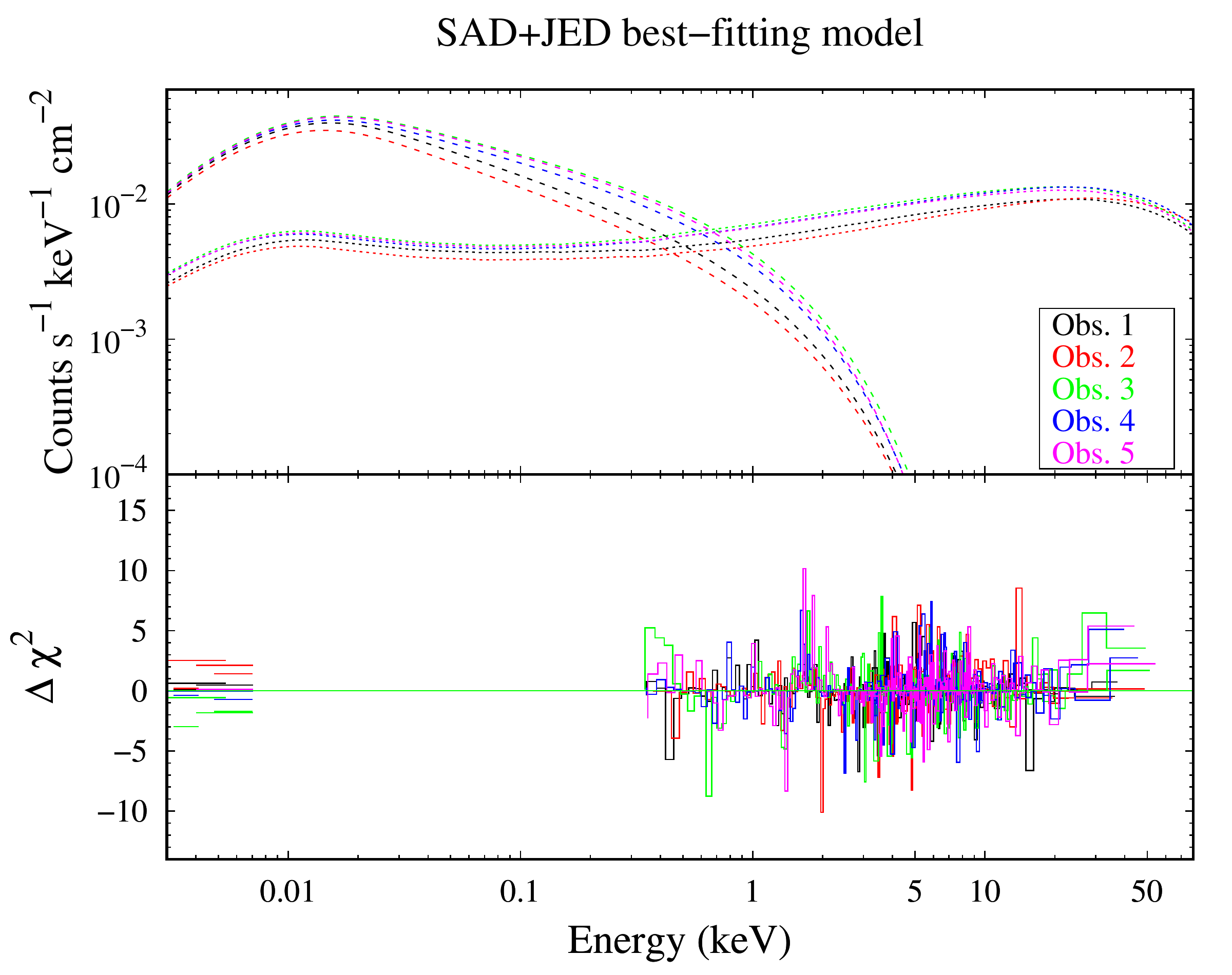}
        \caption{\label{fig:jed} 
                Upper panel: Best-fitting Comptonized SAD and JED model components (dashed and dotted lines, respectively). Lower panel: Residuals of the SAD--JED model.
                }
\end{figure}

\begin{table*} 
                \centering
        \caption{Best-fitting parameters of the SAD--JED model described in Sect. \ref{subsec:jed}: {\sc smallBB+zgauss+highecut*simpl*sad+jed+xillverCp} in \xspec\ notation. 
        The parameters of \xillvercp\ were frozen at the values found in Sect. \ref{subsec:fit} (see Table \ref{params}).
                \label{params_jed}
        }
        \begin{tabular}{l c c c c c c}
                \hline
                &all obs. &obs. 1 & obs. 2& obs.3 &obs. 4& obs. 5\\
                \hline  
                $F_{\textsc{smallBB}}$ ($10^{-12}$ \fluxcgs)& \ser{7.2}{0.6} &&&&& \\ 
                
                \noalign{\medskip}
                $E_{\textsc{zgauss}}$ (keV) &\aer{0.533}{0.009}{0.004}&&&&& \\ 
                $N_{\textsc{zgauss}}$ ($10^{-4}$) &\ser{2.3}{0.4}&&&&& \\                 
                
                \noalign{\medskip}      
                $\Gamma_s$ && \ser{2.55}{0.02} & \ser{2.58}{0.02}  & \ser{2.40}{0.0 }  & \ser{2.44}{0.02} & \ser{2.41}{ 0.02 } \\
                $E_s$ (keV) &&\aer{1.37}{0.16}{0.13}&\ser{1.5}{0.2}&\aer{1.20}{0.08}{0.04}&\aer{1.20}{0.10}{0.09}&\ser{1.13}{0.07} \\
                
                \noalign{\medskip} 
                $\mbh$ (\tento{7} \msun)&\ser{3.69}{0.07}&&&&& \\
                $r_J$ (\rg)&&\ser{ 17.7 }{ 0.5 } 
                & \ser{ 18.6 }{ 0.6 }  & \ser{ 19.1 }{ 0.4 }  & \ser{ 19.5 }{ 0.4 } & \ser{ 18.7 }{ 0.4 } \\
                $\mdot$ ($L\subrm{Edd}/c^2$) && \ser{ 0.79 }{ 0.02 }
                & \ser{ 0.737 }{ 0.014 }  & \ser{ 0.93 }{ 0.02 }  & \ser{ 0.89 }{ 0.02 } & \ser{ 0.90 }{ 0.02 }\\

                        \noalign{\medskip} 
                        $\Gamma_{\textsc{xillverCp}}$ & 1.82(f) &&&&& \\
                        $kT_{\textsc{xillverCp}}$ (keV) & 18.5(f) &&&&& \\
                        $A_{\textrm{Fe,}\textsc{xillverCp}}$ & 2.8(f) &&&&& \\
                        $\log \xi_{\textsc{xillverCp}}$ (\lumcgs\ cm) & 1.71(f) &&&&&\\         
                        $N_{\textsc{xillverCp}}$ (\tento{-5}) & 2.9(f) &&&&& \\
                
                \noalign{\medskip}
                $\Delta \Gamma$ &\ser{0.039}{0.013}&&&&& \\
                
                \noalign{\medskip}
                $\rchisq$ &2052/1999&&&&& \\
                \hline 
        \end{tabular}
\end{table*}

\begin{figure} 
        \includegraphics[width=\columnwidth]{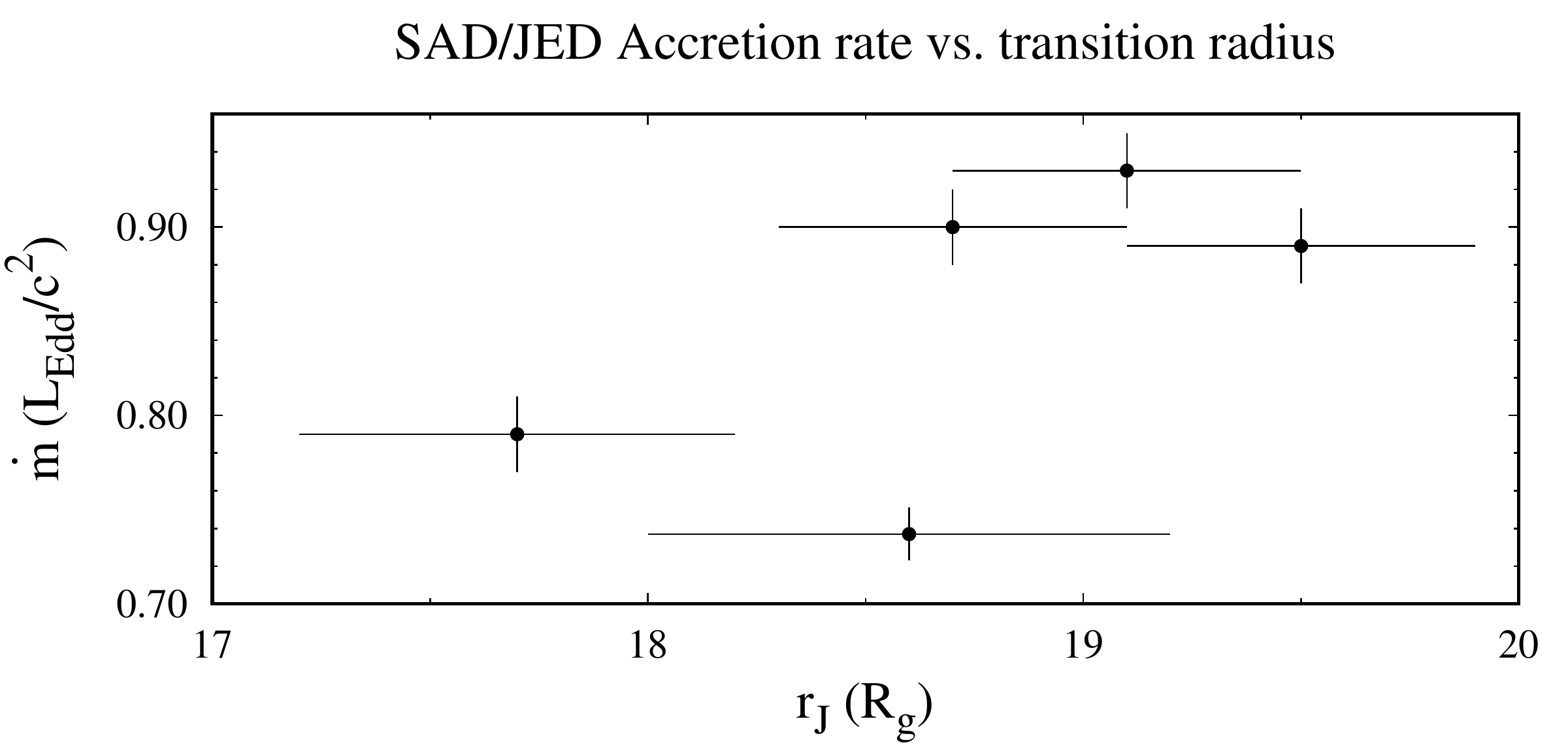}
        \caption{\label{fig:pjed}
                Accretion rate vs. transition radius of the SAD--JED model for the different observations (see Table \ref{params_jed}).}
\end{figure}

\section{Discussion and conclusions}\label{sec:discussion}
The joint \xmm\ and \nus\ campaign on \he\ discussed in this paper allowed us to study the high-energy spectral properties of this unabsorbed Seyfert 1 galaxy in detail, constraining the physics and geometry of its accretion flow. Below we summarize the main results, then we discuss in more detail their physical interpretation.

The source is clearly variable in flux during the campaign; a significant variation is seen between the `low-flux period', corresponding to observations 1 and 2, and the `high-flux period', corresponding to observations 3, 4, and 5. The spectral shape also shows some variability in the soft band (below 10 keV), while little spectral variability is found in the hard band. The data indicate the presence of a correlation between the UV and soft X-ray emission, consistent with a Comptonization origin for the latter, as we discuss below.

The time-averaged spectrum shows a clear indication of a turnover at high energies above 30 keV. This turnover can be phenomenologically reproduced by a power law with exponential cut-off plus a moderate reflection bump; in this case, the cut-off energy is \aer{100}{50}{20} keV. Alternatively, the spectral turnover is nicely described by thermal Comptonization, with an electron temperature of $\sim 20$ keV.

We find the presence of a \fek\ emission line at 6.40 keV (rest-frame), with an intrinsic width of 0.11 keV, and consistent with being constant in flux during our campaign. The line is consistent with originating from a mildly ionized medium and is accompanied by a moderate reflection component, also consistent with being constant. The X-ray spectral properties are consistent with an ionized reflector with $\log \xi \simeq 1.7$ \lumcgs\ cm and an iron overabundance of $\sim 3$. These values explain the moderate broadening of the line. The reflecting material can be identified with the outer part of the accretion disc since we find no strong evidence of relativistic effects due to the proximity of the black hole. 

We confirm the presence of a significant soft X-ray excess below 2 keV in addition to the primary power law. 
Relativistic reflection can reproduce this excess, but only assuming a very low inclination and with a fit of the X-ray spectrum, which is not completely satisfactory, especially in the high-energy band. On the other hand, the broad-band (optical--UV to X-rays) data are consistent with a two-corona scenario. 

\subsection{Two-corona scenario}

According to our results, the warm corona is consistent with having a constant temperature of $\sim 0.5$ keV. However, the observed variability of the photon index of the asymptotic power law implies some physical and/or geometrical variations. We can distinguish between the low-flux period, namely observations 1 and 2 (with $\gammaw \simeq 2.9$) and the high-flux period ($\gammaw \simeq 2.75$). 
The optical depth is consistent with being in the range 16--17 during the low-flux period and in the range 18--20 during the high-flux period. To constrain the geometry of the warm corona, we can estimate the Compton amplification factor $A_w$ (i.e. the ratio between the total power emitted by the corona and power of the seed soft photons from the disc). 
Following the procedure of \cite{cheeses}, 
 later corrected in \cite{pop2019},
we estimate $A_w \simeq 1.1$. Given the values of the photon index, this amplification indicates that the disc is consistent with having an intrinsic emission of around 10\% of the total, rather than being completely passive \cite[see the Appendix in][]{pop2019}. 

The observed anticorrelation between the photon index and the flux of the warm Comptonization component, previously reported in NGC 4593 \citep{4593_2}, indicates that the spectrum of the soft excess hardens as the source brightens. This behaviour could be an effect of the X-ray illumination of the warm corona by the hot one. As the hot corona brightens, the warm corona is more illuminated and thus heated, producing a harder spectrum. 

The parameters of the hot corona do not exhibit a strong variability: the temperature is consistent with 15--20 keV, while the optical depth is around 4 (assuming a spherical geometry). We also estimate an amplification factor $A_h \simeq 13-17$ for the hot corona, with no clear trend between the low-flux and high-flux periods. In any case, the estimate of $A_h$ allows us to estimate the geometrical parameter $g$ that describes the compactness or patchiness of the corona, since $g \simeq 2/A_h$ \citep{cheeses}. We find $g \simeq 0.12-0.15$, indicating that the hot corona intercepts around 12--15\% of the seed soft photons. 
The observed correlation between the primary flux and the flux of the soft excess suggests an interplay between the hot and warm coronae, and can be explained if the photons Comptonized in the hot corona are emitted by the warm corona.
In the spectral model used in this work, the two components are independent. Exploring the consequences of their coupling will be a future extension of the two-corona scenario. We note, however, that the warm corona emits most of the photons in the optical--UV band, similarly to a standard disc. Therefore, our assumption of a hot corona illuminated by a multicolour disc black body can be considered as a fair approximation. 

Finally, we note that the warm Comptonization model for the soft excess has been critically examined by \cite{garcia2019}. These authors argued that in a warm corona the photoelectric opacity is expected to dominate over the Thomson opacity, yielding significant absorption features in the soft X-ray band that are not actually observed. 
\cite{pop2019} addressed this problem by performing new simulations of spectra emitted by warm and optically thick coronae. Petrucci et al. used the radiative
transfer code \textsc{titan} \citep{dumont2003} coupled with the Monte Carlo code \textsc{noar} \citep{dumont2000}, the latter fully accounting for Compton scattering of continuum and lines.
        These simulations show, in a large part of the parameter space, 
        that the warm corona is dominated by Compton cooling and the emitted spectrum presents no 
        strong absorption or emission lines. Furthermore, the spectrum 
        is consistent with the generally observed properties of the soft excess. 
        The results rely on the crucial assumption that the warm corona has a source 
        of internal heating power. In other words, the upper layer of the disc must be heated via dissipation of accretion power, which is possible for example by means of magnetic fields \citep[e.g.][]{gronk2019}\footnote{ This assumption is especially realistic in the SAD--JED configuration, as the SAD portion is threaded by a large-scale vertical magnetic field. Moreover, the existence of a large-scale magnetic field does not preclude the existence of small-scale fields, such as those invoked in earlier works to explain the X-ray emission from accretion disc coronae \citep[e.g.][]{galeev1979}. }.
        This is consistent with the concept of an energetically dominant warm corona covering a quasi-passive disc. The results of these simulations validate warm Comptonization as a scenario to explain the soft excess.

\subsection{A jet-emitting disc?}
The SAD--JED model also
provides a nice description of the data. Perhaps the most striking feature of this model is the relatively small number of free parameters needed to fit the data. There are   essentially two parameters, namely the accretion rate and the SAD--JED transition radius. The accretion rate (in Eddington units) is found to vary between $\sim 0.7-0.8$ in the low-flux period and $\sim 0.9$ in the high-flux period. The data also indicate small ($\sim 10\%$) fluctuations of the transition radius around 19 \rg. Moreover, the SAD model nicely describes the optical--UV emission, both in terms of flux and temperature.
The black hole mass is tightly constrained by the total luminosity and by the observed spectral shape in the optical--UV band. In the SAD--JED model, the precise value of the black hole mass depends on the distance and, potentially, on the other fixed parameters. However, there is good agreement between the best-fitting mass and the independent estimates based on the X-ray variability and the H $\beta$ line.
We note that the Comptonized tail of the SAD is, in this context, essentially a phenomenological component to account for the soft excess. Future extensions of the model will be needed to treat the emission from the warm corona in a more physical and self-consistent way. 

The relation between the radio power of the disc-driven jet and the SAD--JED physical properties is \citep[eq. 3 in][]{jed4}

\begin{equation}
\frac{\nu L_{\nu}}{L\subrm{Edd}} = \tilde{f}_R \, \mdot^{17/12} \, r\subrm{in} \, (r_J-r\subrm{in})^{5/6}
,\end{equation}where $\nu L_{\nu}$ is the radio power and $\tilde{f}_R$ is a dimensionless factor. In general, the radio emission of radio-quiet sources can have different origins \citep[and references therein]{panessa2019}; in any case, assuming that the observed 1.4 GHz flux of \he\ is due to a jet, and taking $\mdot=0.8$ and $r_J=19$ \rg, we derive $\tilde{f}_R=\expom{1.3}{-9}$. 
Interestingly, this factor is not too far from that derived by \cite{jed4} for the X-ray binary GX~339-4 ($\tilde{f}_R = 4.5\times10^{-10}$). High-resolution radio interferometric observations of \he\ will be needed to probe the presence of a radio jet in this source and, potentially, to study its relation with the high-energy emission.

\begin{figure}[]
        \includegraphics[width=1.\linewidth]{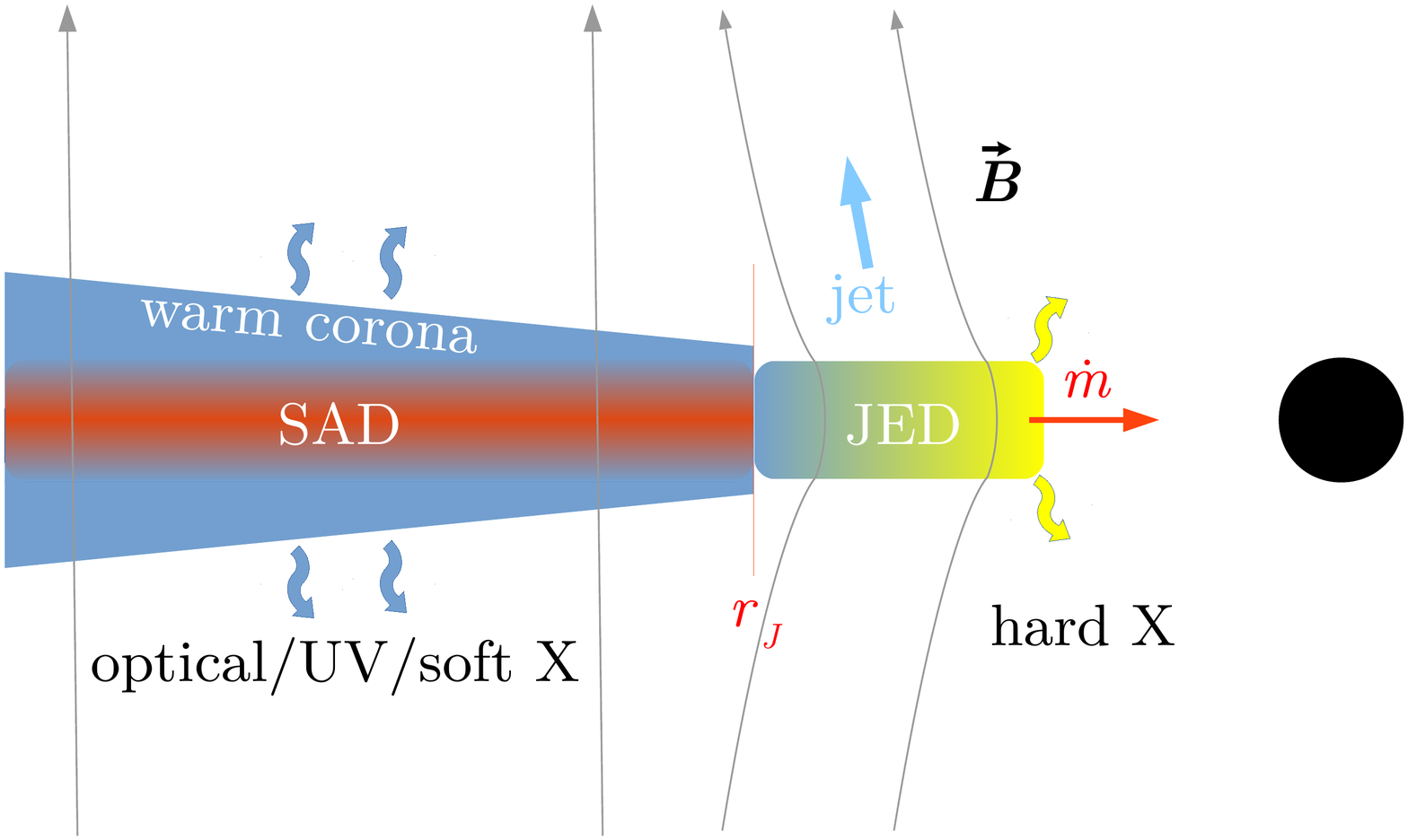}
        \caption{Sketch of the two-corona scenario in which the JED plays the role of the hot corona. The best-fitting parameters for \he\ are $r_J \sim 19$ \rg\ and $\mdot \sim 0.7-0.9 \, L\subrm{Edd}/c^2$.}
        \label{fig:sketch}
\end{figure}

All in all, our results suggest the following tentative scenario (see also the sketch in Fig. \ref{fig:sketch}). The outer part of the accretion flow can be described by a thin standard disc, with a warm upper layer in which most of the gravitational power is released \citep{rozanska2015,cheeses}. This warm corona is responsible for the optical--UV emission and the soft X-ray excess via thermal Comptonization. Below $\sim 20$ gravitational radii, the accretion flow inflates and switches to an inner slim disc corresponding to the hot corona and illuminated by the outer thin disc. 
The flux variability, which is significant on a timescale of a few days, is driven by the variability of the accretion rate. The hot corona in turn illuminates the warm corona, possibly producing more heating (i.e. a harder warm corona spectrum) as the flux increases.

\begin{acknowledgements}
We thank the referee Javier Garc\'ia for detailed comments that significantly improved the manuscript.
F.U. acknowledges financial support from ASI and INAF under INTEGRAL ``accordo ASI/INAF 2013-025-R1''. P.-O.P. and J.F. acknowledge financial support from the CNES French agency and CNRS PNHE. S.B. acknowledges financial support from ASI under grant ASI-INAF I/037/12/0. S.B., A.D.R. and G.P. acknowledge financial contribution from the agreement ASI-INAF n.2017-14-H.O. 
B.D.M. acknowledges support from the European Union's Horizon 2020 research and innovation programme under the Marie Sklodowska-Curie grant agreement No 798726.
G. Marcel acknowledges the \textsc{matplotlib} library \citep{hunter2007} for the production of figures in this paper.

\end{acknowledgements}

   \bibliographystyle{aa} % style aa.bst
   \bibliography{mybib.bib} % your references Yourfile.bib

\appendix

\section{Spectral discrepancy between pn and \nus}\label{app:pn-nustar}

To systematically check the spectral discrepancy between \xmm/pn and \nus, we studied the shape of the spectra from the different observations in the common bandpass 3--10 keV. For this analysis, we also included \xmm/MOS2 data. We did not use the MOS1 data because they are affected by a known hot column issue, which is particularly severe in Timing mode\footnote{\url{https://www.cosmos.esa.int/web/xmm-newton/sas-watchout-mos1-timing}}. We extracted the MOS2 spectra 
from rectangular regions having a width of 40 pixels for the source (coordinate RAWX in the range 284--324) and 26 pixels for the background (RAWX in the range 256--284). The MOS2 spectra were grouped to have at least 30 counts per bin and without oversampling the spectral resolution by a factor greater than 3.

\begin{figure}
        \includegraphics[width=1.\linewidth]{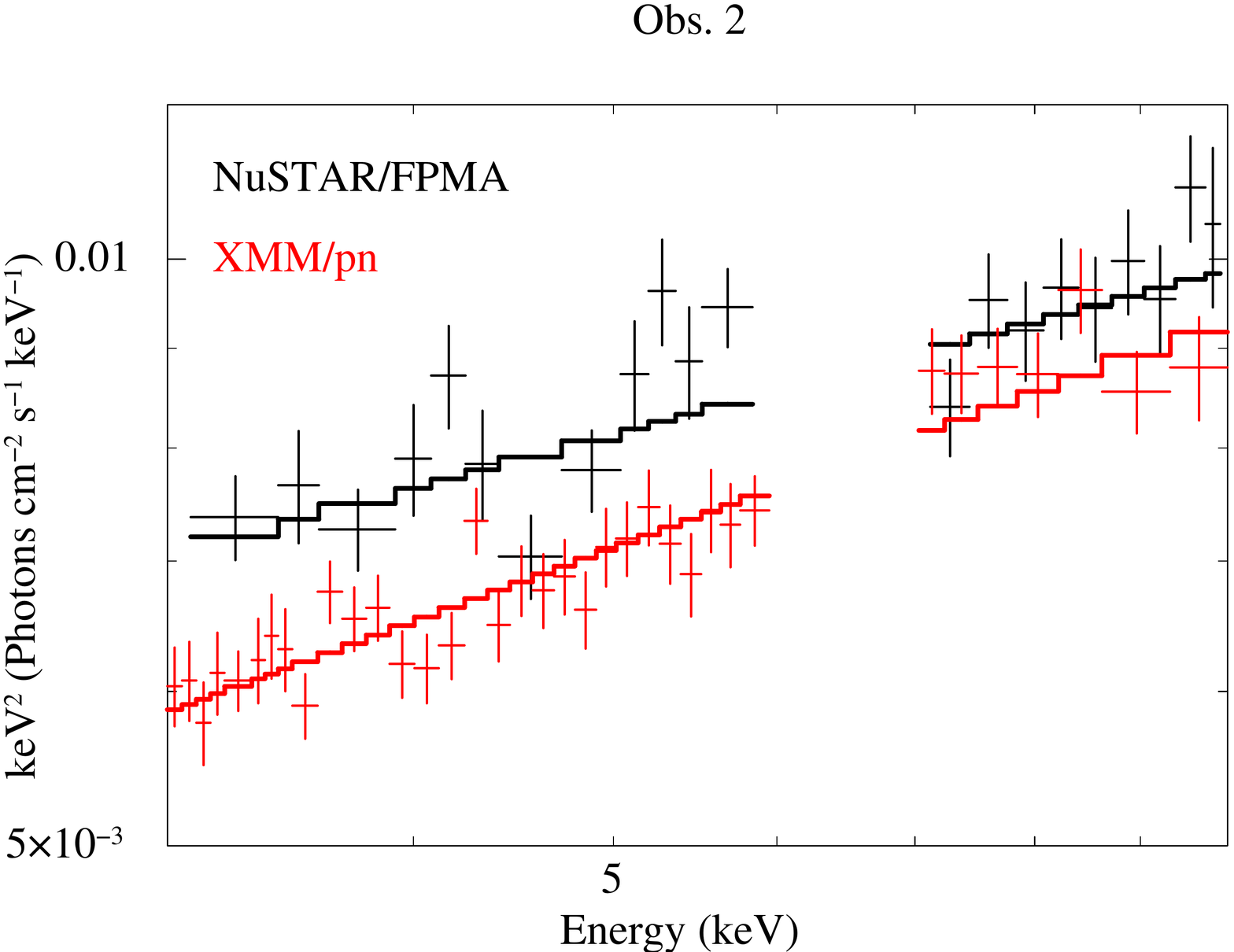}
        \caption{
                Unfolded \nus/FPMA  (black) and \xmm/pn (red) spectra with best-fitting power law for observation 2. 
                }
        \label{fig:discrep}
\end{figure}

We fitted the \nus, pn, and MOS2 spectra with a simple power law in the 3--10 keV band, excluding the 6--7 keV range to avoid the contribution of the \fek\ line. The photon index was left free to vary for each instrument, being tied only between \nus/FPMA  and FPMB. We list in Table \ref{tab:gammas} the best-fitting photon indices. In general, pn yields lower values than \nus. The discrepancy is significant in three observations out of five (observations 2, 4, and 5), whereas it is marginally consistent with zero in the other two (observations 1 and 3). The largest discrepancy is found in observation 2 (see Fig. \ref{fig:discrep}) However, the difference in photon index is always consistent with the average value of $\Delta \Gamma \simeq 0.07$, with an uncertainty of 0.02 from simple error propagation. Concerning the fluxes, pn yields values smaller by $\sim 10\%$ compared with \nus.  
For MOS2, given the poorer signal-to-noise ratio, the photon index has a relatively large uncertainty and is roughly consistent with both \nus\ and pn in the first four observations, while in observation 5 it is consistent only with \nus. Therefore, even though we cannot derive strong conclusions, the MOS2 spectral shape seems to be in slightly better agreement with \nus\ than with pn. On the other hand, the MOS2 fluxes are always smaller than the pn and \nus\ values. This is possibly due to instrumental calibration issues of MOS in Timing mode.

        Finally, we checked whether variable absorption could help to explain the discrepancy between pn and \nus. We first fitted the pn and \nus\ data set (not including MOS2 in this case) in the 3--10 keV range leaving the photon index of the power law free to vary between different observations, but keeping it tied between pn and \nus. We included a free flux cross-calibration constant. Only Galactic column density was included at this stage. We obtained a fit with $\rchisq=793/780$. Then we included an absorption component (\textsc{phabs}) free to vary among the different observations. We obtained only a marginal improvement, namely $\rchisq=786/775$ ($\dchi/\ddof=-7/-5$ and p-value of 0.22 from an F-test). Finally, we left the photon index free to vary between pn and \nus. We found a significant improvement, with $\rchisq=767/770$ ($\dchi/\ddof=-19/-5$; p-value of 0.002 from an F-test) and a column density in addition to the Galactic value always consistent with zero. 

We conclude that the pn-\nus\ spectral discrepancy is likely due to cross-calibration issues between the two instruments. The magnitude of the discrepancy is consistent with being constant among the observations of our campaign. However, since different values have been reported in the literature, there is currently no indication of a systematic discrepancy for which an aprioristic correction is possible. Nevertheless, it is possible to obtain acceptable fits using a multiplicative correction factor, as we did in this work following \cite{ingram2017}.

\begin{table*}
        \begin{center}
                \caption{
                        3--10 keV photon indices ($\Gamma$) and fluxes ($F$) for each observation, as measured by the different instruments. The fluxes are in units of \tento{-11} \fluxcgs.
                        \label{tab:gammas}}
                \begin{tabular}{l c c c c c}
                        \hline 
                         &Obs. 1&Obs. 2&Obs. 3&Obs. 4&Obs. 5\\
                        \hline
                         $\Gamma\subrm{NuS}$ &\ser{1.75}{0.04}&\ser{1.72}{0.04}&\ser{1.77}{0.04}&\ser{1.78}{0.04}&\ser{1.77}{0.04} \\ 
                         $\Gamma\subrm{pn}$ &\ser{1.70}{0.03}&\ser{1.61}{0.04}&\ser{1.73}{0.03}&\ser{1.72}{0.03}&\ser{1.70}{0.03} \\      
                         $\Gamma\subrm{MOS2}$ &\ser{1.73}{0.09}&\ser{1.56}{0.19}&\ser{1.77}{0.09}&\ser{1.81}{0.09}&\ser{1.87}{0.09} \\                               
                        \hline
                        $F\subrm{NuS}$ &\ser{1.71}{0.02}&\ser{1.62}{0.02}&\ser{2.14}{0.02}&\ser{2.09}{0.03}&\ser{2.03}{0.03}\\
                        $F\subrm{pn}$&\ser{1.58}{0.02}&\ser{1.43}{0.02}&\ser{2.03}{0.02}&\ser{1.92}{0.03}&\ser{1.89}{0.02}\\
                        $F\subrm{MOS2}$&\ser{1.42}{0.05}&\ser{1.3}{0.1}&\ser{1.89}{0.06}&\ser{1.72}{0.06}&\ser{1.70}{0.05}\\
                        \hline
                \end{tabular}
        \end{center}
\end{table*}

\section{The RGS spectrum}\label{app:rgs}
From the 2004 \xmm/RGS and pn data, \cite{cardaci2011} detected one \ovii\ emission line in the soft X-ray band, and found no evidence of warm absorption. 
We used the RGS data obtained during our campaign to search for putative emission lines. 
The RGS spectra from different epochs show significant flux variability, but with a modest spectral variability in the continuum. Therefore, to obtain a better signal-to-noise ratio, we co-added the different spectra (separately for the two detectors RGS1 and RGS2). We fitted the co-added spectra in the 0.3--2 keV band. 

First, we fitted the spectra with a simple power law, finding $\rcash=5786/5371$. The fit left significant, positive residuals around 22 \AA\ (i.e. at the energies corresponding to the K$\alpha$ triplet of \ovii).
We then performed
a local fit at 22 \AA, on an interval $\sim 100$ channels wide. Because of the limited bandwidth, we fixed the photon index of the underlying continuum at 2. We detected three significant emission
lines (at the 90\% confidence level), that can be identified as the
resonance (1s$^2$ $^1$S$_0$--1s2p $^1$P$_1$),
intercombination (1s$^2$ $^1$S$_0$--1s2p $^1$P$_{2,1}$), and 
forbidden (1s$^2$ $^1$S$_0$--1s2s $^3$S$_1$) components of the \ovii\ triplet.
Including these lines in the fit over the 0.3--2 keV band, we
found $\rcash=5750/5365$ ($\dcash/\ddof=-36/-6$) without significant residuals attributable to strong atomic transitions. The properties of the triplet are summarized in Table \ref{tab:ovii}.

\begin{table}
        \caption{Emission lines detected in the RGS spectra. $\lambda_T$ and $E_T$ are the theoretical wavelength and energy of the lines (rest-frame), as reported in the \textsc{atomdb} data base (Foster et al. 2012). The flux is in units of \tento{-5} photons cm$^{-2}$ s$^{-1}$. \label{tab:ovii}}
        \centering
        \begin{tabular}{ c c c c c} 
                \hline \hline 
                Line id. & $\lambda_T$ (\AA) &  $E_T$ (keV) & $E_{obs}$ (keV) & Flux \\ 
                \hline 
                \ovii\ (r) & 21.602  & 0.574  & \aer{0.5730}{0.0003}{0.0010} & \aer{2.1}{1.7}{0.9} \\[2pt] 
                \ovii\ (f) & 22.101  & 0.561  & \aer{0.5605}{0.0007}{0.0005} & \aer{2.5}{1.3}{1.2} \\[2pt]  
                \ovii\ (i) & 22.807  & 0.569  & \ser{0.5687}{0.0007} & \aer{2.5}{1.5}{1.2} \\[2pt]  
                \hline          
        \end{tabular}
\end{table}

\section{Physical properties of the SAD--JED solution}\label{app:jed}

The combination of an outer standard accretion disc \citep[SAD,][]{ss1973} and an inner jet-emitting disc \citep[JED,][]{ferreira1997} allows us to reproduce all the typical high-energy spectra of stellar-mass black hole X-ray binaries \citep{jed2}. 
However, the equations involved are self-similar, implying that this model can be extended to AGNs. 

Using the code developed in \cite{jed2,jed3}, we solved for the thermal structure and self-consistently computed the resulting spectra for different values of black hole masses, 
accretion rates, and transition radii between the two different flows. The other parameters of the accretion flow are frozen to those that best reproduce high-luminosity hard states. 
We also assume here that the black hole is not spinning ($a=0$, i.e. $r\subrm{in}=6$ \rg; \citealp{gravitation}). 
The table is calculated in the following ranges of free parameters:
\begin{itemize}
        \item Black hole mass: $m \in [10^6,~ 10^9]$ in solar masses;
        \item Inner disc accretion rate: $\dot{m} \in [10^{-2},~10^2]$ in Eddington units $(\dot{M}\subrm{Edd} \equiv L\subrm{Edd}/c^2)$;
        \item Transition radius: $r_J \in [r\subrm{in}=6,~10^2]$ in gravitational radii.
\end{itemize}

\begin{figure}
        \includegraphics[width=1.\linewidth]{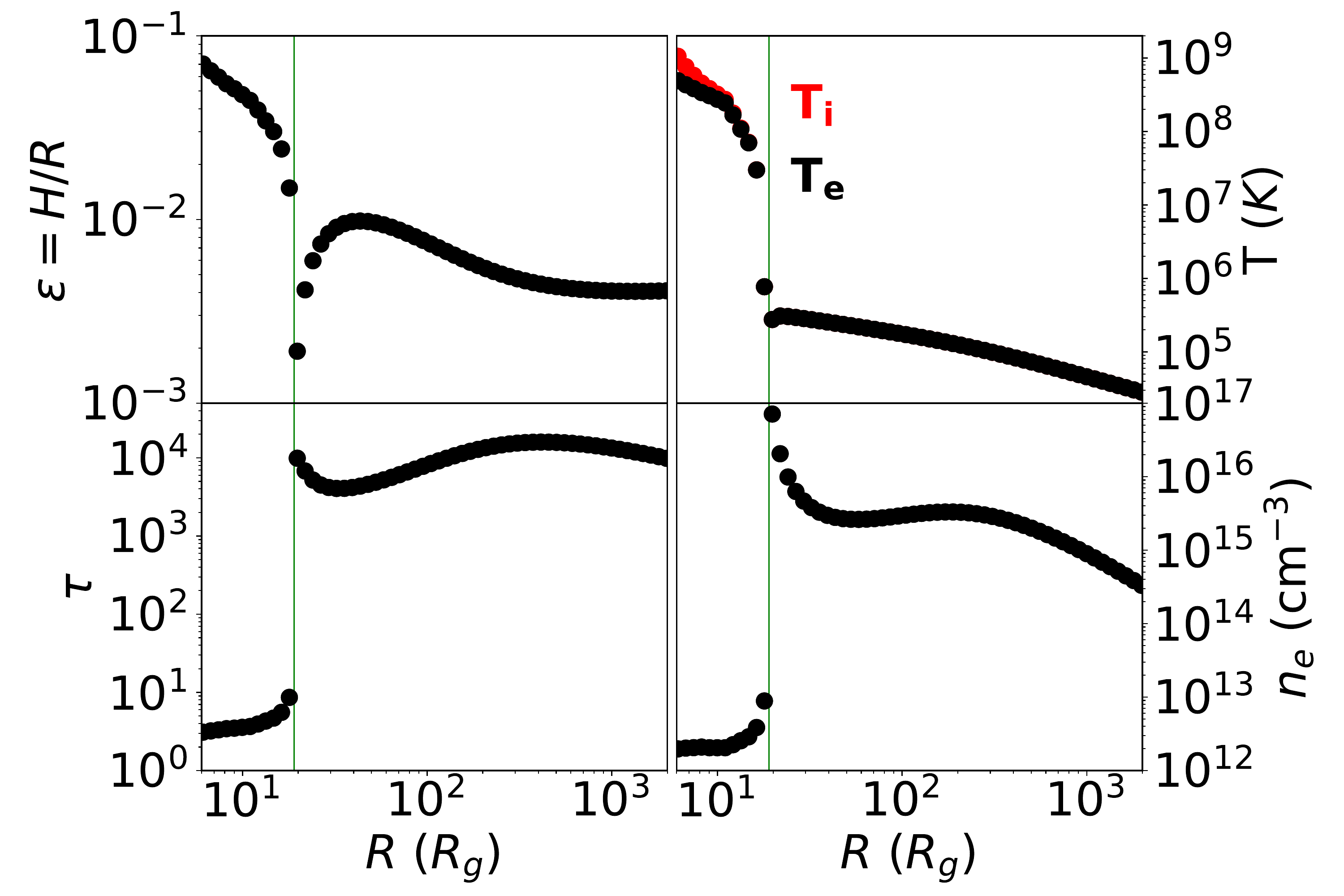}
        \caption{SAD--JED physical structure as a function of the radius. A green vertical line marks the SAD--JED transition at $r_J = 19$.
                Top left: Aspect ratio $\varepsilon = h/r$. Top right: Electron temperature $T_e$ (black) and ion temperature $T_i$ (red) in K. Bottom left: Thomson optical depth $\tau_{T}$. Bottom right: Electron--ion density $n_e=n_i$ in cm$^{-3}$.}
        \label{fig:a1}
\end{figure}

Here we show the physical properties of the SAD--JED solution assuming $r_J=19$ and $\dot{m}_{in}=0.8$ (i.e. average values from the fits; see Sect. \ref{subsec:jed}). In Fig.~\ref{fig:a1}, we show some local properties of the flow: aspect ratio, Thomson optical depth, temperature, and density. 
For a radius greater than $r_J$, the disc is assumed to be a typical $\alpha$-disc (SAD) with $\alpha=0.1$. For the given accretion rate, the SAD is dense, geometrically thin ($\varepsilon < 0.01$) and optically thick ($\tau_T \sim 10^4$); ions and electrons are thermalized and cold ($T_e \sim 10^5$~K).
Below $r_J$, the flow is magnetized and jets are produced (JED). The presence of jets produces a magnetic torque within the disc that accelerates accretion up to supersonic speeds, with Mach number $m_s > 1$. Since the disc density is linked to the Mach number $n_s \propto m_s^{-1}$ \citep{pop2010}, the JED solution has a relatively low density and is geometrically slim or thick ($\varepsilon \sim 0.01-0.1$). The Thomson optical depth is $\sim 3$ and ions and electrons are not thermalized ($T_i \gtrsim T_e$), with $T_e \sim 10^8-10^9$~K. 

\begin{figure}
        \includegraphics[width=1.\linewidth]{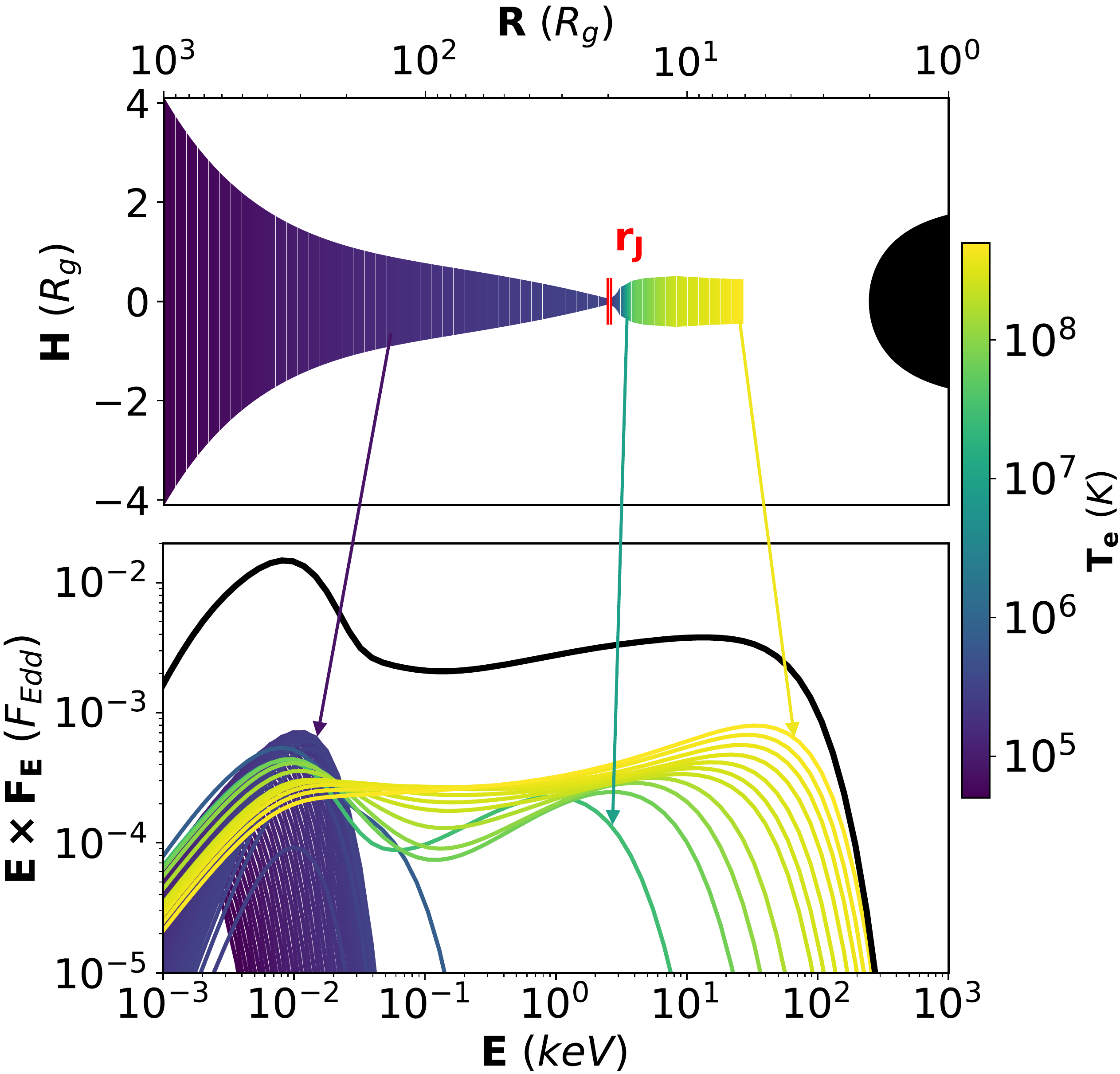}
        \caption{Top panel: Computed geometrical shape of the SAD--JED, divided in annuli, colour-coded according to the electron temperature. 
        Bottom panel: Total emitted spectrum (in black) with the contribution from each annulus, using the same colour-coding as above. The arrows associate three annuli at different radii with their spectrum.}
        \label{fig:a2}
\end{figure}

In Fig.~\ref{fig:a2}, we show the computed geometrical shape of the SAD--JED and its emitted spectrum. The SAD produces multicolour disc black-body spectra, with typical temperatures of $\sim 10$ eV, while the inner JED, below $\sim 10$ \rg, strongly radiates in the hard X-rays. The sum of the different contributions yields a smooth power-law spectrum \citep[see also][]{jed3}. 

\end{document}